\documentclass[useAMS,usenatbib]{mn2e}

\usepackage{hyperref}
\usepackage{graphicx}
\usepackage{mathrsfs} 
\usepackage{graphicx}
\usepackage{color}
\usepackage{amsmath}
\usepackage{amssymb}
\usepackage{varioref}
\usepackage{amssymb}
\usepackage{verbatim}
\usepackage{comment}
\usepackage[normalem]{ulem}
\usepackage{bm} 

\title[Constants of motion in stationary axisymmetric gravitational fields]{Constants of motion in stationary axisymmetric gravitational fields}
\author[C. Markakis]{C. Markakis$^{1,2}$\thanks{E-mail:
charalampos.markakis@uni-jena.de}\\
$^{1}$Department of Physics, University of Wisconsin-Milwaukee, P.O.~Box 413,
Milwaukee, WI 53201, USA\\
$^{2}$Theoretical Physics Institute,  University of  Jena,  07743 Jena, Germany
}
\begin{document}


\pagerange{\pageref{firstpage}--\pageref{lastpage}} \pubyear{2012}

\maketitle

\label{firstpage}

\begin{abstract}
The motion of test particles in stationary axisymmetric gravitational fields is generally nonintegrable unless a nontrivial constant of motion, in addition to energy and angular momentum along the symmetry axis, exists. The Carter constant in Kerr-de Sitter spacetime is the only example known to date. Proposed astrophysical tests of the black-hole no-hair theorem have often involved integrable gravitational fields more general than the Kerr family, but the existence of such fields has been a matter of debate. To elucidate this problem, we treat its Newtonian analogue by systematically searching for nontrivial constants of motion polynomial in the momenta and obtain two theorems. 

First, solving a set of quadratic integrability conditions, we establish the existence and uniqueness of the family of stationary axisymmetric potentials admitting a quadratic constant. As in Kerr-de Sitter spacetime, the mass moments of this class satisfy a ``no-hair'' recursion relation $M_{2l+2}=a^2 M_{2l}$, and the constant is Noether-related to a second-order
 Killing-St\"ackel tensor. Second, solving a new set of quartic integrability conditions, we establish nonexistence of quartic constants. Remarkably, a subset of these conditions is satisfied when the mass moments obey a generalized ``no-hair'' recursion relation $M_{2l+4}=(a^2+b^2)M_{2l+2}-a^2b^2 M_{2l}$. The full set of quartic integrability conditions, however, cannot be satisfied nontrivially by any stationary axisymmetric vacuum potential.

\end{abstract}

\begin{keywords}
gravitation -- black hole physics -- celestial mechanics -- chaos -- stellar dynamics -- galaxies: star clusters  \end{keywords}

\section{Introduction}

Since the work of Euler, Lagrange and Jacobi, the motion of a test particle in a Newtonian dipole field\footnote{Following  terminology in \cite{Misner1973}, a Newtonian dipole will refer to  a pair of fixed positive-mass  centers each of which creates a Newtonian gravitational
   field} has been known to be completely integrable in terms of quadratures. In addition to energy and angular momentum about the symmetry axis, there exists a third nontrivial constant of motion quadratic in the momenta. The  constant was first discovered by  \cite{Euler1760,Euler1764} for two-dimensional (meridional) motion, and the problem is known as the \textit{Euler problem} \citep*{Lukyanov2005a}.
\cite{Lagrange1766} extended Euler's solution to three-dimensional motion and made a further generalization by allowing a Hookian (spring) center to be included  between the two Newtonian (gravitational) centers. This generalization is known as the \textit{Lagrange problem} (cf. \citealt{Lukyanov2005a}).
The problem was further studied by  \cite{Jacobi2009a} using 
separation of variables of the Hamilton-Jacobi equation in prolate spheroidal coordinates.

The literature regarding the problem of two fixed centers is surveyed by  \cite{Lukyanov2005a} and the orbits are studied in \cite{OMathuna2008}.
The quadratic\footnote{A quadratic  constant will  refer to a constant  quadratic in the momenta} constant has been studied by several authors using the Hamilton-Jacobi approach 
\citep{Stackel1890,Whittaker1989,Eddington1915,Kuzmin1956,Lynden-Bell1962,DeZeeuw1985c,DeZeeuw1985b,DeZeeuw1985a}.  If one  considers the distance between the two centers to be imaginary, then one obtains a potential  separable in oblate spheroidal coordinates. This is known in satellite geodesy as the \textit{Vinti potential} and has been used to approximate the gravitational field around the oblate earth \citep{Vinti1960,Vinti1963,Vinti1969,Vinti1971,Vinti1998}. A further generalization is possible by relaxing equatorial plane symmetry. In the oblate case, this   is accomplished  by considering 
 two Newtonian fixed centers with complex-conjugated masses  located at constant
imaginary distance. The resulting \textit{Darboux--Gredeaks potential}
has  been used to  approximate the gravitational field around other oblate planets (\citealt*{Aksenov1963}, \citealt{Lukyanov2005a}).

 \cite{Lynden-Bell2003} has provided a simple and elegant derivation of the quadratic constant in the Euler problem, by noting that  the kinetic part of the constant is the dot product of the angular momenta about the two fixed centers of attraction. He also provided a generalization of the constant for   a class of potentials that satisfy a certain integrability condition. It will be demonstrated in section \ref{sec:cosmoconstantkerrdesitter} that this class is 
essentially that of the Lagrange problem, that is, it  amounts to the addition of a  Hooke term to the potential.

    \cite{Israel1970} and  \cite{Keres1967} have demonstrated that the dipole field of the Euler problem can be regarded as  the Newtonian analogue of the Kerr solution in general relativity. Following Misner's suggestion to seek analogues of the quadratic  constant in Newtonian dipole fields, Carter 
discovered his quadratic constant of motion in  Kerr spacetime by separation of variables \citep*{Carter1968,Carter1977,Misner1973}.
The analogy has been further elucidated by  \cite{Lynden-Bell2003},  \cite{Flanagan2007} and  \cite{Will2009}.

 \cite{Carter1968,Carter2009,Carter2010} also discovered  a generalization of the Kerr solution with a non-zero cosmological constant, describing a rotating black hole in four dimensional de Sitter (or anti de Sitter) backgrounds.  Carter's quadratic constant of motion exists for this class of spacetimes as well   and  has been used to solve  for the orbits in terms of hypergeometric functions  \citep{Kraniotis2004,Kraniotis2005,Kraniotis2011}.
It will be shown in section \ref{sec:cosmoconstantkerrdesitter} that the quadratic constant of the Lagrange two-center problem  
is the Newtonian analogue of the Carter constant in a Kerr-de Sitter
spacetime.

The main motivation behind the study of integrability in stationary axisymmetric gravitational fields is to astrophysically test  the Kerr solution and the no-hair theorem for black holes.
The Kerr solution owes its integrability to
the existence of the Carter constant, which is very useful for analyzing the orbit of a small black hole around a massive black hole. Detecting gravitational waves from these extreme mass ratio inspirals is a prime goal of the proposed   Laser Interferometer Space Antenna, LISA/eLISA \citep{OliverJennrich,Amaro2012a,Amaro2012}.
The influence of gravitational radiation reaction on the evolution of the Carter constant has been used to obtain the  gravitational waveforms from orbits around Kerr black holes. As demonstrated by  \cite{Ryan1995} (see also \citealt{Sotiriou2004}), by measuring the mass,  spin and at least one
more non-trivial moment of the gravitational field of a black hole candidate via gravitational wave observations, one can test the validity of the no-hair theorem. Such tests have been proposed for the supermassive black hole at the centre of the Milky Way, Sgr A*, as reviewed, for example, by \cite{Johannsen2011a}. Work by 
\cite{Glampedakis2006,Gair2008,Psaltis2009a,Johannsen2010b,Johannsen2010a,Johannsen2011b,Psaltis2011,Psaltis2012,Collins2004,Hughes2006,Dubovsky2007,Vigeland2010,Vigeland2010a,Vigeland2011,Johannsen2013}
is a fraction of the vast literature.

In order to provide a systematic framework for testing the Kerr solution, a number of the above articles have introduced ``bumpy'' black-hole spacetimes, which are stationary, axisymmetric, vacuum spacetimes with arbitrary multipole moments (deviating from those of Kerr), in terms of which the stellar orbits and their associated observables are parametrized. This parametrization allows one to astrophysically test the relationship between the multipole moments, and thus probe the validity of the Kerr solution in general relativity. Because the arbitrariness of the moments leads to loss of the Carter constant, some authors have sought other stationary axisymmetric vacuum spacetimes that may be integrable, i.e. that admit a generalized Carter constant
\citep{Brink2008,Brink2008a,Brink2010,Brink2010a,Brink2011}.
 However, there exist conflicting claims in the literature about whether such spacetimes actually exist \citep{Mirshekari2010,Kruglikov2011,Lukes2012}.
 
The subject of the present article is a  systematic approach towards  resolving such conflicts and elucidating the relation between the no-hair theorem and integrability of motion in the Newtonian regime. We use a direct approach \citep{Hietarinta1987} to systematically search for a nontrivial constant  polynomial in the momenta. This approach consists of directly solving the Killing equations and certain integrability conditions. In section \ref{sec:QuadraticInvariantsKT} we   show that
 the constant in the Euler and Lagrange two-center problems is the \textit{unique quadratic constant}  for motion around a stationary axisymmetric massive object with equatorial reflection symmetry. 
Although this result is implicit in other work (cf. \citealt{Kalnins2009,Kalnins2010,Kalnins2012})  no  explicit
  proof of uniqueness\footnote{The advantage of the derivation in \cite{Will2009}, based on  a multipole expansion,   is that it establishes uniqueness among all   stationary  axisymmetric potentials, but this uniqueness is restricted to  invariants  constructed from a  certain combination of  the linear or angular momentum and position vectors.} 
existed in the literature prior to that of  \cite{Will2009}.
The merits of the  systematic procedure outlined here are that it provides a   complete proof (as it involves no assumptions regarding separability or the form of  the quadratic constant) and that it is generalizable to more complicated systems with higher-order Killing tensors, as illustrated  in subsequent sections.
In section \ref{sec:GeneralizedNoetherSym}  the quadratic constant is shown to be Noether-related.
In section \ref{sec:QuarticInvariantsKT}  we consider the next
natural generalization, a constant \textit{quartic} in the momenta. Using the direct approach, we show that \textit{no such constant exists} for stationary axisymmetric vacuum potentials  in Newtonian gravity.

Extrapolating these uniqueness and non-existence theorems  into the relativistic regime provides highly suggestive evidence in favour of the conjecture that a stationary axisymmetric vacuum spacetime is integrable if and only if it belongs to the Kerr (or Kerr-de Sitter) class. Besides the theoretical implications, this  is important from an onservational/astronomical point of view. For example, this conjecture is a working assumption of a number of articles discussing astrophysical tests of the Kerr black hole solution 
\citep{Apostolatos2009,Lukes2010,Lukes2010a,CONTOPOULOS2011}.
Conclusions, and a discussion of this matter,  are given in section \ref{sec:summary}.



\section{invariants quadratic in the momenta}
\label{sec:QuadraticInvariantsKT}

\subsection{Killing equation and integrability conditions}
\label{sec:Killingequationandintegrabilityconditions}

The motion of a test particle in a Newtonian gravitational field $\Phi$  is independent of the particle mass, so for simplicity we can set the latter equal to unity. This motion is described by an
action functional
\begin{equation}  \label{eq:Action3D}
S[x,p]=\int_{t_1}^{t_2} dt[p_i \dot x^i-H(x,p)]
\end{equation}
where position $x^i$ and momentum $p_i$ are treated as independent variables,
\begin{equation}  \label{eq:Hamiltonian3D}
H(x,p)=\frac{1}{2}g^{ij} (x)p_ip_j+\Phi (x)
\end{equation}
is the Hamiltonian, 
$g^{ij}$ is the inverse of the Euclidian metric $g_{ij}$    in $\mathbb{\mathbb{E}}^3$   and summation over repeated indices is implied. Indices are raised and lowered with this metric throughout the paper. Unless otherwise noted,  we will be using Cartesian coordinates, so that 
$g_{ij}=\delta_{ij}$
where $\delta_{ij}$ is the Kronecker delta.
The Cartesian components of  the canonical momentum $p_i=g_{ij}\dot{x}^j$ will then  coincide with those of the kinetic momentum $\dot x^i$.

Spatial translations and rotations in $\mathbb{\mathbb{E}}^3$ are generated by the vector fields 
\begin{equation}  \label{eq:XiRi}
\bm{X}_k={\bm e_k}, \qquad \bm{R}_i=\varepsilon^{k}_{\phantom{0}ji}x^j \bm{e}_k
\end{equation}
where ${\bm e_k}=\{\bm{\hat x},\bm{\hat y},\bm{\hat z}\}$ are Cartesian basis vectors and $\varepsilon_{ijk}$ is the Levi-Civita tensor. We are interested in orbits of a test particle in the vicinity of a \textit{stationary} and \textit{axisymmetric} massive object, an object whose gravitational potential satisfies
\begin{align} 
\partial_t\Phi&=0  \label{eq:PhiStationary}\\
\pounds_{\bm{\varphi}}\Phi&={\varphi ^k}{\partial _k}{\Phi}=(x\partial_y-y\partial_x)\Phi=0 \label{eq:PhiAxisymmetric}
\end{align}
Here, $\pounds_{\bm{\varphi}}$ denotes the Lie-derivative along the Killing vector
field\begin{equation}  \label{eq:AxialZeta}
\bm{\varphi} \equiv \bm{R}_z=x \,\bm{\hat y}-y\, \bm{\hat x} 
\end{equation}
 that generates rotations about the  $z$-axis (the axis of symmetry). 
By virtue of Noether's theorem,  invariance with respect to time translations and azimuthal rotations  implies respectively conservation of the Hamiltonian $H$ and the component
\begin{equation}  \label{eq:Lz}
L_z= \varphi^i p_i=R_z^i \,p_i=x\, p_y-y\, p_x
\end{equation}
of  angular momentum.
Rotations about the $x$ and $y$
axes, generated  by
the vector fields
$\bm{R}_x$
and
$\bm{R}_y$
are \textit{not}  considered symmetries of the problem, so the  components
$L_x= R_{x}^i\, p_i$ and $L_y=R_{y}^i\, p_i$
of  angular momentum are \textit{not} generally conserved. For example, the potential around a rotating massive object (such as a star or planet) in hydrostationary equilibrium,  may be
expected to be axisymmetric, but not spherically symmetric, due to the rotationally induced deformation.  
Without further symmetries, $H$ and $L_z$ are the only independent integrals of    motion and, since
the motion of test particles is three-dimensional, the problem is in general non-integrable. 
Nevertheless, one may seek special types of stationary axisymmetric potentials that admit a third nontrivial constant of motion and are therefore integrable.

 If one seeks a constant of motion linear in the momenta, then one is quickly led to  rotational or translational symmetries as the only choices, which have been already exhausted as explained above. The next natural step is to seek  potentials admitting a nontrivial constant of motion  \textit{quadratic} in the momenta,
\begin{equation}  \label{eq:QuadraticInvariant}
I(x,p)=K^{ij} (x)p_i p_j+K(x),
\end{equation}
where the symmetric tensor $K^{ij}$ and the scalar $K$ are functions of position.
The  above quantity is conserved iff it commutes with the Hamiltonian, in the sense of a vanishing Poisson bracket:
\begin{equation}  \label{eq:PoissonBracket}
\frac{dI}{dt}=\{I,H\}\equiv\frac{\partial I}{\partial x^k}\frac{\partial H}{\partial p_k}
-\frac{\partial I}{\partial p_k}\frac{\partial H}{\partial x^k}=0.
\end{equation}
Substituting the Hamiltonian \eqref{eq:Hamiltonian3D} and the ansatz \eqref{eq:QuadraticInvariant} into the above Poisson bracket yields
\begin{equation}  \label{eq:PoissonBracketQuadratic}
\{I,H\}=\partial^k K^{ij}p_ip_jp_k+( \partial^jK-2K^{jk}\partial_k\Phi)p_{j}
\end{equation}
where $\partial_k=\frac{\partial }{\partial x^{k}}$. In order that this Poisson bracket vanish for all orbits, the following necessary and sufficient conditions \citep{Boccaletti2003} must be satisfied:
\begin{align}  \label{eq:KillingEquationsQuad}
\partial^{(k}K^{ij)}&=0 \\
\partial^j K &= 2 K^{jk}\partial_k\Phi,  \label{eq:KillingConditionQuad}
\end{align}
where parentheses  denote symmetrization over the enclosed indices; that is, $A^{(ij)}=\frac{1}{2}(A^{ij}+A^{ji})$.
Eq. \eqref{eq:KillingEquationsQuad} is a Killing equation in Euclidian space; a solution  $K^{ij}$ to the above equations will be referred to as a Killing-St\"ackel tensor \citep{Carter1977}.
From Eq. \eqref{eq:KillingConditionQuad}, we obtain a necessary, but not sufficient, integrability condition
\begin{equation}  \label{eq:IntegrabilityCondQuad}
\partial ^{[i} (K^{j]k}\partial_k\Phi)=\frac{1}{2}\partial^{[i}\partial^{j]} K=0
\end{equation}
where square brackets denote antisymmetrization over the enclosed indices, that is $A^{[ij]}=\frac{1}{2}(A^{ij}-A^{ji})$.
 
Our assumptions of stationarity and axisymmetry already guarantee the existence of two independent solutions to the above set of equations. First, the metric itself is  already a Killing-St\"ackel
tensor, because equations
\eqref{eq:KillingEquationsQuad}--\eqref{eq:IntegrabilityCondQuad} are  satisfied by   $K^{ij}=\frac{1}{2}g^{ij}=\frac{1}{2}\delta^{ij}$, $K=\Phi$, and the associated conserved quantity is simply the Hamiltonian \eqref{eq:Hamiltonian3D}. Secondly, the above equations are also satisfied by the reducible Killing-St\"ackel
tensor 
$K^{ij}=\varphi^{i}\varphi^{j}$, with $K=0$, implying conservation of the quantity $L_z^2$.
This second case is  reducible to the linear invariant \eqref{eq:Lz}
associated with the axial Killing vector $\varphi^i$. 
We now explore the possibility of a third independent solution that can render the problem integrable.

The  conditions \eqref{eq:KillingEquationsQuad}--\eqref{eq:IntegrabilityCondQuad} suggest a systematic method \citep{Hietarinta1987} for obtaining integrable potentials and the associated invariants of motion:

\begin{enumerate}
  \item The  tensor $K^{ij}$ may be computed by solving Eq. \eqref{eq:KillingEquationsQuad} subject to the symmetries of the problem.   \item With this tensor known, the integrability condition \eqref{eq:IntegrabilityCondQuad} provides a key  restriction on the  Newtonian potential $\Phi$. One may solve  this   condition (e.g. via a multipole method)  to obtain  a family of  integrable potentials.  \item Given a family of potentials $\Phi$ that satisfy this integrability condition,  one may obtain the scalar function $K$ by  integrating the components of Eq. \eqref{eq:KillingConditionQuad}.
\end{enumerate}

With this method as the basis of our analysis, we proceed to carry out the prescribed steps in more detail.

\subsection{Rank-two Killing-St\"ackel tensors} 

One may straightforwardly solve Eq. \eqref{eq:KillingEquationsQuad} by noticing that
the solution
must be polynomial in the Cartesian coordinates. This is easily seen in one or two dimensions (c.f. Appendix \ref{ch:appendixInhomKilling}) and can be generalized to arbitrary dimensions and tensor rank \citep{HORWOOD2008}. In three dimensions, Eq.  \eqref{eq:KillingEquationsQuad} constitutes an overdetermined system of ten equations
for the six independent components of the symmetric tensor  $K^{ij}$. As shown by   \cite{HORWOOD2008}, the most general solution  to this system
is a sum of symmetrized products of the translational and rotational  vectors \eqref{eq:XiRi}:  
\begin{equation}  \label{eq:KijQuadGeneralcoordfree}
{\bm K} 
={K^{ij}} \bm{X}_i\!\otimes\! \bm{X}_j= {\mathcal{A}^{ij}\bm{X}_i\!\otimes\! \bm{X}_j} + 2{\mathcal{B}^{ij}} \bm{X}_i\!\otimes\! \bm{R}_j
+ {\mathcal{C}^{ij}}\bm{R}_i\!\otimes\! \bm{R}_j
\end{equation}
The components ${K^{ij}}$ of the tensor ${\bm K}$ are polynomial of second order in the Cartesian coordinates $x^i$ and can be written as 
\begin{equation}  \label{eq:KijQuadGeneral}
{K^{ij}} = {\mathcal{A}^{ij}} + 2{\mathcal{B}^{k(i}}{\varepsilon^{j)}}_{kl}{x^l} 
+ {\mathcal{C}^{mn}}{\varepsilon^i}_{km}\varepsilon{^j}_{ln}{x^k}{x^l}
\end{equation}
where $\mathcal{A}^{ij}=\mathcal{A}^{(ij)},\,\mathcal{C}^{ij}=\mathcal{C}^{(ij)}$ are \textit{symmetric} $3\times3$ constant  matrices and $\mathcal{B}^{ij}$ is a \textit{nonsymmetric} $3\times3$ constant matrix. While  a superficial counting results in 21 independent coefficients, the actual number of independent coefficients is  20. This is because the main diagonal elements of  $\mathcal{B}^{ij}$ appear in $K^{ij}$ only in the combinations 
$\mathcal{B}^{xx}-\mathcal{B}^{yy}$, $\mathcal{B}^{yy}-\mathcal{B}^{zz}$, $\mathcal{B}^{zz}-\mathcal{B}^{xx}$ and the sum of these three terms is zero.

\subsection{Isometries}
 One may considerably reduce the number of unknown coefficients by imposing known symmetries of the action functional $S$  on the orbital invariant $I$.
\begin{enumerate}
\item
 \textit{Stationarity} corresponds to invariance of $S$ under the  group action of  $(\mathbb{R},+)$ which represents time translations
  $t\rightarrow t+\delta t$. 
  This symmetry has already been taken into account, since all quantities have no explicit dependence on time $t$  and an additive term $\partial I/\partial t$
has been set to zero in Eq. \eqref{eq:PoissonBracket}. 

\item
\textit{Axisymmetry} corresponds to
invariance of $S$ under the group action of SO(2) which represents infinitesimal rotations 
$\{x,y,z\} \rightarrow \{x+y\,\delta\varphi , y-x\,\delta\varphi,z\}$ about
the $z$-axis.
This symmetry is generated by the Killing vector \eqref{eq:AxialZeta} and may be imposed by requiring
\begin{equation}  \label{eq:AxisymmetryI}
\left( x\frac{\partial}{\partial y}-y\frac{\partial}{\partial x}
+p_x\frac{\partial}{\partial p_y}-p_y\frac{\partial}{\partial p_x} \right)
I=0
\end{equation}
 With
$I$ given by Eq. \eqref{eq:QuadraticInvariant}, the above condition can be shown  to be equivalent to the requirement that the functions $K^{ij}$ and $K$ remain unchanged under rotations about the $z$-axis: 
\begin{align}  \label{eq:AxisymmetryKij}
\pounds_{\bm \varphi}K^{ij}&={\varphi ^k}{\partial _k}{K^{ij}} - {K^{kj}}{\partial _k}{\varphi ^i} - {K^{ki}}{\partial _k}{\varphi ^j}=0 \phantom{aaa}\\
\pounds_{\bm \varphi}K&={\varphi ^k}{\partial _k}{K}=0 \label{eq:AxisymmetryK}
\end{align}
Evidently, the symmetries 
\eqref{eq:AxisymmetryI}, \eqref{eq:AxisymmetryKij}, \eqref{eq:AxisymmetryK} are inherited from the Hamiltonian, which is seen by the fact that these equations are also satisfied by replacing $I, K^{ij}, K$ with $H,\frac{1}{2}\gamma^{ij},\Phi$ in the above three equations respectively. Substituting the general solution \eqref{eq:KijQuadGeneral}
into Eq. \eqref{eq:AxisymmetryKij} yields the axisymmetry constraints
\begin{align}  \label{eq:AxisymConstraints1}
\mathcal{A}^{xy}&=\mathcal{A}^{xz}=\mathcal{A}^{yz}=\mathcal{A}^{xx}-\mathcal{A}^{yy}=0\\ \label{eq:AxisymConstraints2}
\mathcal{B}^{xz}&=\mathcal{B}^{zx}=\mathcal{B}^{yz}=\mathcal{B}^{zy} \nonumber \\ 
&=\mathcal{B}^{xx}-\mathcal{B}^{yy}=\mathcal{B}^{xy}+\mathcal{B}^{yx}=0\\ \label{eq:AxisymConstraints3}
\mathcal{C}^{xy}&=\mathcal{C}^{xz}=\mathcal{C}^{yz}=\mathcal{C}^{xx}-\mathcal{C}^{yy}=0
\end{align}

\item Further simplification is possible by assuming  \textit{equatorial plane reflection symmetry}. This corresponds to invariance of 
$S$ under the discrete group action of 
$\mathbb{Z}_2$ which represents reflections  $\{x,y,z,p_x,p_y,p_z\}\rightarrow \{x,y,-z,p_x,p_y,-p_z\} $ 
about the equatorial plane.
(This assumption is not necessary for integrability, c.f. \cite{Lynden-Bell2003}, but we retain it for simplicity).
Imposing this symmetry on the invariant  \eqref{eq:QuadraticInvariant} 
 leads to the constraints
\begin{align}  \label{eq:EquatorConstraints1}
\mathcal{A}^{xz}&=\mathcal{A}^{yz}=0\\  \label{eq:EquatorConstraints2}
\mathcal{B}^{xx}&=\mathcal{B}^{yy}=\mathcal{B}^{zz}\\  \label{eq:EquatorConstraints3}
\mathcal{B}^{xy}&=\mathcal{B}^{yx}=0  \\  \label{eq:EquatorConstraints4}
\mathcal{C}^{xz}&=\mathcal{C}^{yz}=0
\end{align}
\end{enumerate}

 Note that the constraints \eqref{eq:AxisymConstraints1}--\eqref{eq:AxisymConstraints3} and \eqref{eq:EquatorConstraints1}--\eqref{eq:EquatorConstraints4} are independent,    since axisymmetry and reflection symmetry are separate assumptions. Imposing these  two sets of constraints on the tensor \eqref{eq:KijQuadGeneral}
yields
\begin{align}  \label{eq:KijAxisymReflCompon}
{\bm K}
&=\mathcal{A}^{yy}(\bm{X}_x\!\otimes \!\bm{X}_x+\bm{X}_y\!\otimes \!\bm{X}_y)+\mathcal{A}^{zz}\bm{X}_z\!\otimes\! \bm{X}_z
\nonumber\\
&+\mathcal{C}^{yy}(\bm{R}_x\!\otimes\! \bm{R}_x+\bm{R}_y\!\otimes\! \bm{R}_y)+\mathcal{C}^{zz}\bm{R}_z\!\otimes\! \bm{R}_z
\nonumber\\
&= {{\cal C}^{yy}}
\left( {\begin{array}{*{20}{c}}
{{y^2} + {z^2}}&{ - xy}&{ - xz}\\
{ - xy}&{{x^2} + {z^2}}&{ - yz}\\
{ - xz}&{ - yz}&{{x^2} + {y^2}}
\end{array}} \right) \phantom{00000000} \nonumber\\
&  +(\mathcal{A}^{zz}-\mathcal{A}^{yy})
   \left( {\begin{array}{*{20}{c}}
{{0}}&0&0\\
0&{{0}}&0\\
0&0&{{1}}
\end{array}} \right)+{\cal A}^{yy} \left( {\begin{array}{*{20}{c}}
{{1}}&0&0\\
0&{{1}}&0\\
0&0&{{1}}
\end{array}} \right) 
 \nonumber\\ 
 &+(\mathcal{C}^{zz} - \mathcal{C}^{yy})
\left( {\begin{array}{*{20}{c}}
{{y^2}}&{ - xy}&0\\
{ - xy}&{{x^2}}&0\\
0&0&0
\end{array}} \right) 
\phantom{00000000000}
\end{align}

This is the most general solution to Eq. \eqref{eq:KillingEquationsQuad}
consistent with the symmetry  
$(\mathbb{R},+)\times \text{SO(2)} \times \mathbb{Z}_2$. For  clarity, we set  $\kappa\equiv \mathcal{C}^{yy}$, $\lambda\equiv \mathcal{A}^{zz}$, $\mu\equiv \mathcal{C}^{zz}-\mathcal{C}^{yy}$  and $a\equiv \sqrt{(\mathcal{A}^{zz}-\mathcal{A}^{yy})/\mathcal{C}^{yy}}$. The   parameter $a$ is allowed to be  real or imaginary and will be shown to depend on the gravitational source. (This reparametrization entails no loss of generality, as the sign of $(\mathcal{A}^{zz}-\mathcal{A}^{yy})/\mathcal{C}^{yy}$ is unrestricted.
We have also excluded the possibility of vanishing  $\mathcal{C}^{yy}$
with nonvanishing
$(\mathcal{A}^{zz}-\mathcal{A}^{yy})$.
This case  would lead to a Killing-St\"ackel tensor $(\mathcal{A}^{zz}-\mathcal{A}^{yy})\delta^i_z \delta^j_z$, reducible to a Killing vector $\delta^i_z$  that generates translations  along the $z$-axis. But this is  \textit{not}  a symmetry of the problem by assumption, i.e. we have excluded cylindrical symmetry.)
The above   solution may then be written as
\begin{equation}  \label{eq:KijAxisymReflCoef}
{K^{ij}} = \kappa A^{ij}+ \lambda \ \! g^{ij}+\mu \ \! \varphi^i\varphi^j
\end{equation}
where  $g_{ij}=\delta_{ij}$ is the Euclidian metric, 
$\bm \varphi$  is the axial Killing vector \eqref{eq:AxialZeta},
\begin{align}  \label{eq:KijAxisymReflReduced}
{A^{ij}} 
&= \delta^{kl} R_k^i R_l^j+ a^ia^j-a^2 g^{ij} \nonumber\\
&= (r^2g^{ij}-x^ix^j)-(a^2g^{ij}-
 a^i a^j) \nonumber\\
 &=g^{mn} \varepsilon^i_{\phantom{i}km} \varepsilon^j_{\phantom{j}ln} (x^k x^l-a^k a^l)\nonumber\\
&=g^{mn} \varepsilon^i_{\phantom{i}km} \varepsilon^j_{\phantom{j}ln} 
(x^k+a^k)(x^l-a^l) 
\end{align}
is a new nontrivial Killing-St\"ackel tensor,   $r=\sqrt{x^k x_k}$  is a radial coordinate and $a^i=a\delta^i_z$ are the components of the vector $\bm a=a\, \bm{ \hat z}$. 
With $\kappa, \lambda,\mu$ regarded as arbitrary coefficients, the general solution \eqref{eq:KijAxisymReflCoef}  is  a linear combination of  three independent solutions: the  known Killing-St\"ackel tensors 
 $\gamma^{ij}=\delta^{ij}$ and $\varphi^i\varphi^j$, associated with stationarity and axisymmetry, and a third independent solution
$A^{ij}$ associated with a Noether symmetry to be discussed later.
The above expression gives the \textit{most general rank-two Killing tensor} in stationary-axisymmetric vacuum potentials. This completes step (i) of the prescribed procedure.

\subsection{Integrability condition: no-hair relation}

Although the solution \eqref{eq:KijAxisymReflReduced} satisfies Eq. \eqref{eq:KillingEquationsQuad} and the symmetries of the problem, it does not  lead to a conserved quantity unless the condition \eqref{eq:KillingConditionQuad} is satisfied. Our next step is thus to
use the integrability condition \eqref{eq:IntegrabilityCondQuad} of Eq. \eqref{eq:KillingConditionQuad} to obtain a family of  potentials $\Phi$ for which  $A^{ij}$ leads to a conserved quantity. The Newtonian gravitational field around an axisymmetric object is completely characterized by a set of mass multipole moments $\{M_L\}$, by means of the expansion 
\begin{equation}  \label{eq:MultipoleExpansion}
\Phi=-\sum_{L=0}^{\infty}\frac{M_L}{r^{L+1}}P_L(z/r) 
\end{equation}
where $P_L$ are the Legendre polynomials. This expansion is consistent with a stationary axisymmetric vacuum potential that vanishes at infinity.

If one also requires equatorial plane reflection symmetry, then the potential $\Phi$ must be an even function of $z$,  so the odd moments $M_1, M_3, \ldots$ 
must vanish. Then, substituting   $K^{ij}=A^{ij}$ and $\Phi$,
as given  by eqs. \eqref{eq:KijAxisymReflReduced} and \eqref{eq:MultipoleExpansion}, into the integrability condition \eqref{eq:IntegrabilityCondQuad}, we find by straightforward algebra that the latter is satisfied 
\textit{if and only if} the even multipole moments satisfy the  recursion relation:
\begin{equation} \label{eq:MomentsRecursionOne}
M_{L+2}=a^2 M_{L}\quad(L=0,2,\ldots)
\end{equation}
or, equivalently,
\begin{equation} \label{eq:MomentsMa}
M_{L}=a^L M_{0}  \quad(L=0,2,\ldots).
\end{equation}
where $a^L$ denotes  the $L$-th power of $a$ in the above two equations. 
This relation is analogous to the ``no-hair'' relation 
for Kerr black holes
\citep{Israel1970,Hansen1974,Will2009}.
All nonzero multipoles are determined
by the mass  $M_0$ and the parameter $a$.  
If $a$ is real (imaginary) then the quadrupole moment  
$M_2=a^2 M_{0}$ is positive (negative) and the expansion \eqref{eq:MultipoleExpansion}
describes  the field of a prolate (oblate)  object. 
Summing the Legendre series
\eqref{eq:MultipoleExpansion}
using Eq. \eqref{eq:MomentsMa} yields \citep{Trahanas2004,Will2009}
\begin{equation} \label{eq:PhiDipole}
\Phi=-\frac{M_{0}/2}{\sqrt{x^2+y^2+(z-a)^2}}
-\frac{M_{0}/2}{\sqrt{x^2+y^2+(z+a)^2}}
\end{equation}
In the prolate case,  the above potential is that of Euler's 
three body problem: a test mass moving in the  gravitational field of two point sources, each of mass $M_{0}/2$, fixed at positions $\pm a \,\hat z$.

 In the oblate case,  the replacement  $a\rightarrow \text{i}\,a$ (where $\text{i}^2=-1$)  allows one to write the above potential  as 
\begin{equation} \label{eq:PhiDipoleComplex}
\Phi=-\frac{M_0 }{\rho+a^2z^{2}/ \rho^3}
\end{equation}
where $\rho$ is an ellipsoidal coordinate defined  by
\begin{equation} \label{eq:ElliposidalCoordinate}
\frac{{{x^2} + {y^2}}}{{{\rho^2} + {a^2}}} + \frac{{{z^2}}}{{{\rho^2}}} = 1
\end{equation}
\cite{Vinti1960,Vinti1963,Vinti1969,Vinti1971,Vinti1998} has shown that the above potential is the most general  solution to the three dimensional Laplace equation that seperates the Hamilton-Jacobi equation in oblate spheroidal coordinates. (A similar statement can be made for the potential    \eqref{eq:PhiDipole} in prolate spheroidal coordinates).
Vinti's potential  has been frequently used in satellite geodesy  to approximate the gravitational field around the oblate earth.

We have so far shown that 
\textit{the solutions \eqref{eq:KijAxisymReflReduced} and \eqref{eq:PhiDipole} are the unique stationary, axisymmetric, equatorially symmetric, vacuum solutions to eqs. \eqref{eq:KillingEquationsQuad} and \eqref{eq:IntegrabilityCondQuad}}.
(Note that we have excluded the possibility of cylindrical symmetry as we are interested in the gravitational field of massive objects with compact support near the origin.) This completes step (ii) of our procedure.

\subsection{Existence and uniqueness of the quadratic invariant}

The last step towards obtaining an invariant is to obtain
the scalar contribution $K$ to the invariant \eqref{eq:QuadraticInvariant}.
This is done by solving the condition  \eqref{eq:KillingConditionQuad} with      $K^{ij}=A^{ij}$ and $\Phi$
  given respectively by eqs. \eqref{eq:KijAxisymReflReduced} and \eqref{eq:MultipoleExpansion}. The solution  $K=A$ is  obtained by   integrating  the   $x$ component of Eq. \eqref{eq:KillingConditionQuad}  with respect to $x$, or the $y$ component with respect to $y$. Up to some additive constant, we find
\begin{align} \label{eq:KDipole}
A&=2\int {dx \,A^{ix}\partial_i \Phi} 
=2\int {dy \,A^{iy}\partial_i \Phi} \nonumber  \\
&=
\frac{M_{0}a(z+a)}{\sqrt{x^2+y^2+(z+a)^2}}
-\frac{M_{0}a(z-a)}{\sqrt{x^2+y^2+(z-a)^2}} 
\end{align}
 Finally, substituting Eq. \eqref{eq:KijAxisymReflReduced} into \eqref{eq:QuadraticInvariant}  and using the Lagrange identity, the quadratic invariant is written
\begin{align} \label{eq:quadinvariantdipole}
I&=A^{ij}p_ip_j+A \nonumber\\
&= r^2|\bm{p}|^{2}-|\bm{x} \cdot \bm{p}|^2-(a^2|\bm{p}|^2 - |\bm a \cdot \bm{p}|^2
 )+A \nonumber\\
&= |\bm x \times \bm p|^2 -|\bm a \times \bm{p}|^2 +A \nonumber\\
&=[(\bm x+\bm a)\times  \bm p] \cdot [(\bm x-\bm a)\times  \bm p]
 +A 
\end{align}
with $A$  given by Eq. 
\eqref{eq:KDipole}.
The above expression holds for the prolate case,   $a=|a|$.  
In the oblate case, $a=\text{i}\,|a|$, the above expression  still gives a true result.
In the spherically symmetric limit, $a\rightarrow0$,  the above invariant reduces to  $I\rightarrow |\bm x \times \bm p|^2$ which is a natural consequence of conserved net angular momentum. The first integrals $H, L_z, I$ are independent and in involution (that is, $\{H,L_z\}=0, \{H,I\}=0$ and $\{I,L_z\}=0$). Thus, the system is Liouville-integrable.

We have shown that the potential \eqref{eq:PhiDipoleComplex} is the  unique stationary, axisymmetric, equatorially symmetric, vacuum
potential admitting a nontrivial constant of motion quadratic in the momenta,  Eq. \eqref{eq:quadinvariantdipole}. The potential \eqref{eq:PhiDipoleComplex} can be considered  the Newtonian analogue \citep{Keres1967,Israel1970,Lynden-Bell2003,Will2009}
of the Kerr solution in general relativity. The analogy is  manifest when the Kerr metric is written in Kerr-Schild coordinates \citep{Kerr}. The associated constant of motion is traditionally obtained by separating the Hamilton-Jacobi equation in Boyer-Lindquist coordinates \citep{Carter1968,Misner1973}. It was in this way that Carter originally discovered his constant, following Misner's suggestion to seek analogies to the constant \eqref{eq:quadinvariantdipole} in Newtonian dipole fields.
This analogy  persists in the presence of a cosmological constant as demonstrated  in the following section.

Another interesting property  is the following: One may interchange  the roles of the metric $g_{ij}$ and potential $\Phi$  with those of the Killing-St\"ackel tensor $A_{ij}$ and scalar field $A$ respectively. Then the  momentum map is generated by   $I=A^{ij}p_ip_j+A$ (which plays the role of the Hamiltonian) and the  quadratic constant of motion is  given by $H= \frac{1}{2} g^{ij}p_ip_j+\Phi$. This duality also exists in the relativistic case of the Kerr spacetime \citep{CarterPC}.

\subsection{Cosmological constant: analogy with a Kerr-de Sitter spacetime}
\label{sec:cosmoconstantkerrdesitter}
One may generalize the result of the previous section in various directions, by relaxing certain assumptions. The  assumption that the motion is in vacuum may be relaxed by adding terms with a non-vanishing Laplacian to the multipole expansion \eqref{eq:MultipoleExpansion} of the gravitational potential. For example, one may attempt to add a spherically symmetric power-law potential of the form $r^l$. Doing so, and repeating the previous steps,  leads to the same expression \eqref{eq:KijAxisymReflReduced} for the Killing tensor $A^{ij}$ as before. Direct substitution shows that the integrability condition \eqref{eq:IntegrabilityCondQuad} is satisfied \textit{if and only if} the   moments $\{M_L\}$ satisfy the  no-hair relation \eqref{eq:MomentsRecursionOne}  and
the only power law term that can be added is  proportional to $r^2$.
This leads to a potential
\begin{align} \label{eq:PhiDipoleLamda}
\Phi=&-\frac{M_{0}/2}{\sqrt{x^2+y^2+(z+a)^2}}
-\frac{M_{0}/2}{\sqrt{x^2+y^2+(z-a)^2}}
\nonumber\\&-\Lambda (x^2+y^2+z^2)
\end{align}
In the prolate case, with $a$   real,
the above potential is that of two force centers, with Newton's inverse square law of  gravitational attraction supplemented by a linear cosmological constant  (or Hooke) term.  (Note that the last term in the above equation may be regarded  as the contribution of a  spring  connecting the test particle to the origin or as  the contribution of two springs connecting the test particle to the two  centers   
at $\pm a \, \bm{\hat z}$).
The associated constant of motion 
is given by Eq. \eqref{eq:quadinvariantdipole} with
\begin{align} \label{eq:ADipoleLamda}
A=& 
\frac{M_{0}a(z+a)}{\sqrt{x^2+y^2+(z+a)^2}}
-\frac{M_{0}a(z-a)}{\sqrt{x^2+y^2+(z-a)^2}} 
\nonumber\\&+2\Lambda a^2 (x^2+y^2)
\end{align}
In the oblate case, with $a$   imaginary, the last term may again be interpreted as the contribution of a cosmological constant. With the substitution $a\rightarrow \text{i}\,a$, the  potential \eqref{eq:PhiDipoleLamda}  can be naturally regarded as the Newtonian analogue of a Kerr-de Sitter spacetime and 
the constant \eqref{eq:ADipoleLamda} as the  analogue of the Carter constant in that spacetime \citep{Carter1968}. For this spacetime the metric has a $tt$ component given by  $g_{tt}=-(1+2 \Phi)$ in Kerr-Schild coordinates, with $\Phi$  given by Eq. \eqref{eq:PhiDipoleLamda} and $a$ replaced by $\text{i}\, a $.

  \cite{Lynden-Bell2003} provided a  generalization of the quadratic constant 
\eqref{eq:quadinvariantdipole} and the two-center potential \eqref{eq:PhiDipole}. His result holds for potentials satisfying a certain    integrability condition, which takes the form of a wave equation, $(\partial^2/\partial r_1^2-\partial^2/\partial r_2^2)(r_1 r_2 \Phi)=0$ in his two-center  coordinates $\bm r_{1,2}=\bm r \pm\bm a$ (\citealt{Lynden-Bell2003}, section 3). 
Assuming equatorial symmetry and substituting  the multipole expansion \eqref{eq:MultipoleExpansion} into this integrability condition, we recover the no-hair relation \eqref{eq:MultipoleExpansion} and  the  potential \eqref{eq:PhiDipole}. If we attempt to add power law terms of the form $r_1^l, r_2^l$ to the potential $\Phi$ we find that the only possibility is
$r_1^2+r_2^2=2(r^2+a^2)$, giving rise to the potential
 \eqref{eq:PhiDipoleLamda}
up to a constant. We infer that Lynden-Bell's generalization accounts for the presence of  a cosmological constant term, which gives rise to  the Newtonian analogue of the Kerr-de Sitter spacetime  discussed above.

The  direct approach employed here is  more algorithmic and computationally intensive compared to other  approaches  \citep{Lynden-Bell2003,Will2009}. But the advantages of the direct approach are that it establishes uniqueness and that it can be straightforwardly generalized to higher order or relativistic invariants. We have shown that  the unique stationary, axisymmetric, equatorially symmetric potential admitting an invariant quadratic in the momenta is given by Eq. \eqref{eq:PhiDipoleLamda}.
The assumption of equatorial symmetry may be relaxed \citep{Lynden-Bell2003} by dropping  conditions \eqref{eq:EquatorConstraints1}-\eqref{eq:EquatorConstraints4}, but we do not pursue this here.
 
\subsection{  Generalized Noether symmetry}\label{sec:GeneralizedNoetherSym}

The generalized Noether theorem may be stated as follows \citep{Ioannou2004}. Consider the $\epsilon$-family of infinitesimal transformations
\begin{equation} \label{eq:phasespacetranformations}
x^{i}\rightarrow x^{i}_\epsilon=x^{i}+\epsilon \, K^i(x,\dot x,t)
\end{equation}
which depend on position \textit{and} \textit{velocity}, for a small parameter $\epsilon$.
If this family of transformations leaves the action $S=\int L \,dt$ invariant, or, equivalently, changes the Lagrangian $L$
by a total time derivative
of some scalar $K(x,t)$,
\begin{equation}  \label{eq:Lagrnagianepsilontimederiv}
L\rightarrow L_\epsilon=L-\epsilon \frac{dK}{dt}
\end{equation}
then the quantity 
\begin{equation}  \label{eq:invariantGeneralizedNoether}
I=\frac{\partial L}{\partial \dot x^i}K^i+K 
\end{equation}
is a constant of motion. Since  the family \eqref{eq:phasespacetranformations} of transformations is velocity-dependent, it is not generally considered a family of  diffeomorphisms. Nevertheless,  it is  a generalized symmetry of the action and Noether-related to an invariant of the form \eqref{eq:invariantGeneralizedNoether}.

Conversely, if the quantity  $I$
is a constant of motion, then the
 $\epsilon$-family of transformations generated by 
$K^i(x,\dot x,t)$, obtained by solving the linear system \citep{Ioannou2004}
\begin{equation}  \label{eq:GeneralizedNoetherLinearSystem}
\frac{\partial^2 L}{\partial \dot x^i \partial \dot x^j}K^i=\frac{\partial I}{\partial \dot x^j},
\end{equation}
is a generalized symmetry of the action. 

The motion of a test particle in a Newtonian graviational field can be obtained from the  Lagrangian  
$L(x,\dot x)=\frac{1}{2}g_{ij} \dot x^i \dot x^j +\Phi$. If   the motion admits a   quadratic  invariant $I(x,\dot x)=K_{ij} \dot x^i \dot x^j +K$, then the  inverse Noether theorem \eqref{eq:GeneralizedNoetherLinearSystem} implies that the 
$\epsilon$-family of transformations  \eqref{eq:phasespacetranformations} generated 
by $K^i=K^{ij}\dot x_j$ is 
a  generalized  symmetry of the action. We infer, for the problem of 
the previous section with $K^{ij}$ given by Eq. \eqref{eq:KijAxisymReflCoef},  that the action 
\eqref{eq:Action3D}
is invariant under three  families of infinitesimal transformations:

\begin{enumerate}
\item 
 The  Killing-St\"ackel tensor $\varphi^i \varphi^j$ is Noether-related to the family
 of transformations
\[ x^{i}_\epsilon=x^{i}+\epsilon \, \varphi^i \varphi^j \dot x_j
=x^{i}+\epsilon \, L_z \, \varphi^i \]
The  tensor $\varphi^i \varphi^j$ is of course reducible to the axial Killing vector $\varphi^i$,  related to the family of diffeomorphisms 
$x^{i}_\epsilon=x^{i}+\epsilon \, \varphi^i$. These  represent azimuthal rotations and give rise to conservation of angular momentum \eqref{eq:Lz}.

\item The metric  $g^{ij}$ is a Killing-St\"ackel tensor and is  Noether-related to the family of transformations
\[ x^{i}_\epsilon=x^{i}+\epsilon \, g^{ij} \dot x_j=
x^{i}+\epsilon \, \dot x^{i} \]
or, equivalently, 
\[
x^{i}_\epsilon(t)=x^{i}(t)+\epsilon \, \dot x^{i}(t) =x^{i}(t+\epsilon)
\]
The metric tensor  $g^{ij}$ is therefore Noether--related to invariance with respect to time translations $t \rightarrow t + \epsilon$ and gives rise to consevation of energy (or the Hamiltonian) given by Eq. \eqref{eq:Hamiltonian3D}.

\item
The irreducible Killing-St\"ackel tensor $A^{ij}$ given by eq.
\eqref{eq:KijAxisymReflReduced}
is Noether-related to the family  of transformations
\[
 x^{i}_\epsilon=
 x^{i}+\epsilon \, A^{ij} \dot x_j
\]
or, equivalently, 
\begin{align}
 \bm x_\epsilon
&=
 \bm x+\epsilon \, [(\bm x  \times \dot{\bm x}) \times \bm x 
 -(\bm a  \times \dot{\bm x}) \times \bm a]\nonumber\\
&=
 \bm x+\epsilon \, [(\bm x -\bm a) \times \dot{\bm x}] \times (\bm x + \bm a)
\nonumber
\end{align}
This a-posteriori knowledge of the symmetry transformation allows a fast ``derivation'' of the quadratic invariant via the Noether procedure: varying the action
of the two-center problem  with respect to the above family of transformations yields no change, while  the Lagrangian changes by $-\epsilon \,dA/dt$ with $A$  given by eq.
\eqref{eq:ADipoleLamda}. Then, Eq. \eqref{eq:invariantGeneralizedNoether}   leads immediately to the quadratic invariant
\eqref{eq:quadinvariantdipole}.

As mentioned above, these transformations are a symmetry of the action,   giving rise to the constant of motion \eqref{eq:KDipole},
 but are not diffeomorphisms
since they depend on position and velocity.
Nevertheless, writing $ \epsilon \, \dot x_j(t)=
x^j(t+\epsilon)-x^j(t)$
they can be expressed as transformations in position and time:
\[
 x^{i}_\epsilon (t)=
 x^{i}(t)-A^{ij} x_{j}(t)+ A^{ij} x_j(t+\epsilon)
\]
where $A^{ij}$ is a function of $x^i(t)$.
\end{enumerate}
The Noether theorem and its inverse, expressed
by Eqs.
  \eqref{eq:phasespacetranformations}-\eqref{eq:GeneralizedNoetherLinearSystem},
also hold
for a relativistic Lagrangian and can be used to show that the four constants of geodesic motion in a Kerr (or Kerr-de Sitter) spacetime are also Noether-related to symmetries of the action.  Axisymmetry is related  to conservation of angular momentum about the symmetry axis as in case (i) above.  Case (ii) discussed above  has two analogues in general relativity:  Stationarity (invariance of the Lagrangian under   time translations along the integral curves of a timelike Killing vector)  is related to conservation of energy or Hamiltonian. Metric affinity 
(invariance of the action under proper time translations) is related to conservation of the magnitude of  four-velocity (or the super-Hamiltonian) and is associated with the four-metric being a Killing tensor.
Finally, a family of transformations analogous to those of case (iii) is related to the Carter constant of motion (cf. \citealt{Padmanabhan2010}, page 381).

\section{invariants QUARTIC in the momenta}
\label{sec:QuarticInvariantsKT}

\subsection{Killing equation and integrability conditions}\label{sec:polybound}

In the  limit $a \rightarrow 0$, the rank-2 Killing tensor  \eqref{eq:KijAxisymReflReduced} of the 2-centre potential   reduces to a  combination  $R_x^i R_x^j+R_y^i R_y^j+R_z^i R_z^j$ of the  axial Killing vectors $R_x^i, R_y^i, R_z^i$ of the spherically symmetric 1-centre potential. Intuitively, one might then expect a hierarchy, whereby  a  4-centre potential  (with source centres at $\pm a, \pm b $) admits a rank-4 Killing tensor  (reducible to a combination of rank-2 tensors in the 2-centre limit  $a \rightarrow b$),  an 8-centre potential admits a rank-8 Killing tensor, and so forth. In light of this, we consider the next natural generalization of the previous section, that is, invariants quartic in the momenta
\begin{equation}   \label{eq:QuarticInvariant}
I(x,p)=K^{ijkl}(x)p_i p_j p_k p_l + K^{ij}(x)p_i p_j+K(x)
\end{equation}
associated with a rank-four Killing-St\"ackel tensor $K^{ijkl}$. The  above quantity is conserved iff it commutes with the Hamiltonian, in the sense of a vanishing Poisson bracket, Eq. \eqref{eq:PoissonBracket}.
Evaluating the bracket with $I$ given by Eq. \eqref{eq:QuarticInvariant}
and $H$ given by Eq. \eqref{eq:Hamiltonian3D} yields
\begin{align}  \label{eq:PoissonBracketQuartic}
\{I,H\} &= 
{\partial ^m}{K^{ijkl}}{p_i}{p_j}{p_k}{p_l}{p_m} 
\nonumber\\&+ ({\partial ^k}{K^{ij}} -4{K^{ijkl}}{\partial _l}\Phi ){p_i}{p_j}{p_k} 
\nonumber\\&+ ({\partial ^k}K - 2{K^{ik}}{\partial _i}\Phi ){p_k} 
\end{align}
Demanding strong integrability, that is, requiring that this Poisson bracket vanish for all orbits, we have
\begin{align}  
{\partial ^{(m}}{K^{ijkl)}}&=0 \label{eq:KillingEq4th1} \\
{\partial ^{(k}}{K^{ij)}} &=4{K^{ijkl}}{\partial _l}\Phi \label{eq:KillingEq4th2}  \\
{\partial ^k}K &= 2{K^{ik}}{\partial _i}\Phi   \label{eq:KillingEq4th3}
\end{align}
 As in the previous section, a solution to Eq. \eqref{eq:KillingEq4th3} exists  only if the following integrability condition is satisfied:
\begin{equation}  \label{eq:IntegrabilityCondQuart3}
\partial ^{[i} (K^{j]k}\partial_k\Phi)=0
\end{equation}
which is identical to  condition \eqref{eq:IntegrabilityCondQuad}.
The above set of equations is employed in \cite{Hietarinta1987}
 to find quartic invariants for various two-dimensional systems. 

Eq. \eqref{eq:KillingEq4th2} may be regarded as an inhomogeneous Killing equation and  is also subject to an integrability condition. The general solution to  this equation  and its integrability condition are obtained in Appendix \ref{ch:appendixInhomKilling}.
 We find that a solution to Eq. \eqref{eq:KillingEq4th2} in $\mathbb{E}^2$ exists only if the following integrability condition is satisfied:
\begin{align}  \label{eq:IntegrabilityCondQuart2}
{\partial _{zzz}}({K^{yyyi}}{\partial _i}\Phi)   
 - 3{\partial _{zzy}}({K^{yyzi}}{\partial _i}\Phi ) =
\nonumber\\ {\partial _{yyy}}({K^{zzzi}}{\partial _i}\Phi   ) 
- 3{\partial _{yyz}}({K^{yzzi}}{\partial _i}\Phi   )
\end{align}
where $i$ is summed over the $y$ and $z$ components and
 $\partial_{ij...k}$ is an abbreviation for $\partial_i\partial_j...\partial_k$.
Extending this integrability condition to $\mathbb{E}^n$ is straightforward and, as shown in  Appendix \ref{ch:appendixInhomKilling}, given by
\begin{align}  \label{eq:IntegrabilityCondQuart22}
&\partial_{nml}f_{ijk}-\partial_{nmi}f_{ljk}-\partial_{njl}f_{imk}+\partial_{nji}f_{lmk}
\nonumber\\-&\partial_{kml}f_{ijn}+\partial_{kmi}f_{ljn}+\partial_{kjl}f_{imn}-\partial_{kji}f_{lmn}=0
\end{align}
 where $f^{ijk}=4{K^{ijkl}}{\partial _l}\Phi$. Eq. \eqref{eq:IntegrabilityCondQuart2} follows from \eqref{eq:IntegrabilityCondQuart22} by setting $n=m=i=z$ and $i=j=k=y$.   Eq. \eqref{eq:IntegrabilityCondQuart2} will suffice for our present purposes, as we shall restrict attention to two-dimensional motion in section
 \ref{sec:nohairquartic} and beyond. If, in addition to $H$ and $L_z$, a third constant of motion    exists for all initial conditions in three dimensions, then this constant ought to also be conserved in the special case of  orbits with $L_z=0$ that lie on a meridional plane.  That is, three dimensional motion is integrable only if meridional motion is integrable.\ We may thus set $x=0$ and study motion restricted on\ the meridional ($y$-$z$) plane first. If  such motion is found to be integrable, generalization to three dimensions is straightforward by virtue of  axisymmetry.

Eqs. \eqref{eq:KillingEq4th1}--\eqref{eq:IntegrabilityCondQuart2} suggest a systematic method  for obtaining  potentials admitting quartic invariants:

\begin{enumerate}
\item
The  tensor $K^{ijkl}$ may be computed by solving Eq. \eqref{eq:KillingEq4th1} subject to the symmetries of the problem. 

\item With this tensor known, the integrability condition  \eqref{eq:IntegrabilityCondQuart2} provides a   restriction on the  Newtonian potential $\Phi$. Solving   this   condition   yields  a family of possibly integrable potentials.

\item Given a family of potentials $\Phi$ which satisfy the  condition \eqref{eq:IntegrabilityCondQuart2},  one may obtain the tensor $K^{ij}$ by solving the inhomogenous Killing equation \eqref{eq:KillingEq4th2}. 

\item With $K^{ijkl}$ and $K^{ij}$ known, Eq. \eqref{eq:IntegrabilityCondQuart2} provides a further restriction on the potential $\Phi$. 

\item If  a family  of potentials $\Phi$ are found that satisfy the above integrability conditions,  then one may obtain the scalar function $K$ by  integrating the components of Eq. \eqref{eq:KillingEq4th3}.

\end{enumerate}

We  now proceed to carry out the above steps in more detail. 


\subsection{Rank-four Killing-St\"ackel tensors} \label{rank4KStensors}
In $\mathbb{E}^3$, Eq. \eqref{eq:KillingEq4th1}
 constitutes an overdetermined system of 21 equations for the 15 unknown independent components of the symmetric tensor $K^{ijkl}$. Similarly, in $\mathbb{E}^2$, Eq. \eqref{eq:KillingEq4th1} is a system of 6 equations for 5 unknowns. The system   
 may be solved by noticing that each component of $K^{ijkl}$ must be a fourth order polynomial in the Cartesian coordinates $x^i$.   \cite{HORWOOD2008} considered the  equation $\partial ^{(j}K^{i_1\cdots i_N)}=0$ for a tensor $K^{i_1\cdots i_N}$ of arbitrary valence $N$ and showed that the general solution    
 is given by
\begin{align} \label{eq:KillingGeneralSolution}
\bm K = \sum\limits_{L = 0}^N {\left( {\begin{array}{*{20}{c}}
N\\
L
\end{array}} \right)
\mathcal{C}_{L}^{{i_1} \cdots {i_{L}}{i_{L + 1}} \cdots {i_N}}{\bm X_{{i_1}}} \otimes  \cdots   } 
\nonumber\\
\cdots  \otimes {\bm X_{{i_{L}}}}\otimes {\bm R_{{i_{L + 1}}}} \otimes  \cdots  \otimes {\bm R_{{i_N}}}
\end{align}
where the objects $\mathcal{C}_{ L}^{{i_1} \cdots  {i_{N}}}$, labelled by $L=0,...,N$, are constant and subject to the symmetries $\mathcal{C}_{L}^{{i_1} \cdots {i_{L}}{j_{L + 1}} \cdots {j_N}}=\mathcal{C}_{L}^{{(i_1} \cdots {i_{L}}{)(j_{L + 1}} \cdots {j_N})}$.
The above expression, for $N=4$ and with $\bm X_i$, $\bm R_j$
 given by Eq. \eqref{eq:XiRi}, provides the general solution to Eq. 
\eqref{eq:KillingEq4th1}. 
Imposing known isometries on the quartic invariant 
\eqref{eq:QuarticInvariant}
constrains the coefficients 
 $\mathcal{C}_{ L}^{{i_1} \cdots  {i_{N}}}$

Stationarity has already been imposed (as the system is autonomous). For  three dimensional motion,  axisymmetry may be imposed via the requirement that the above Killing-St\"ackel tensor is Lie-derived by the axial vector $\bm \varphi=\bm R_z$,
that is
\begin{align}
{\pounds_{\bm \varphi} }{K^{ijkl}} &= {\varphi ^m}{\partial _m}{K^{ijkl}} - {K^{mjkl}}{\partial _m}{\varphi ^i} - {K^{mikl}}{\partial _m}{\varphi ^j} \nonumber\\&- {K^{mijl}}{\partial _m}{\varphi ^k} - {K^{mijk}}{\partial _m}{\varphi ^l}=0
\end{align}
(For  purely meridional motion,  enforcing the above condition does not change the relevant coefficients of   $K^{ijkl}$  projected to the meridional plane.) \ In addition, we also require $\mathbb{Z}_2$ reflection symmetry about the equatorial plane, that is, the replacement
 $\{z,p_z\}\rightarrow \{-z,-p_z\} $ 
 leaves the quantity $K^{ijkl}p_ip_jp_kp_l$  unchanged.
 Since the latter quantity is a fourth order polynomial in the (Cartesian) position and momentum variables, imposing the above isometries is straightforward via algebraic manipulation software such as  Mathematica.
Imposing the above symmetries significantly reduces the number of independent polynomial coefficients. 
Then, after dropping   trivial terms such as  $g^{(ij}g^{kl)}$, $g^{(ij}\varphi^k\varphi^{l)}$, $\varphi^i\varphi^j\varphi^k\varphi^l$ which correspond to reducible Killing-St\"ackel tensors, we find that the most general nontrivial solution
to \eqref{eq:KillingEq4th1} subject to the symmetry $(\mathbb{R},+)\times \text{SO(2)} \times \mathbb{Z}_2$ can be written in the remarkably simple form
\begin{equation} \label{eq:KillingTensorRankFour}
{K^{ijkl}} = \kappa A^{(ij}B^{kl)} + \mu g^{(ij} C^{kl)}
\end{equation}
where $A^{ij}$ is given by Eq. 
\eqref{eq:KijAxisymReflReduced} and $B^{ij},C^{ij}$ are given by the same equation with
the  replacements $\bm a \rightarrow \bm b$, $\bm a \rightarrow \bm c$ respetively: 
\begin{align}  \label{eq:KijAxisymReflReducedQart}
{B^{ij}} &= 
g^{mn}\varepsilon^i_{\phantom{i}km} \varepsilon^j_{\phantom{j}ln} (x^k+b^k)(x^l-b^l) \\
{C^{ij}} &=
g^{mn} \varepsilon^i_{\phantom{i}km} \varepsilon^j_{\phantom{j}ln} (x^k+c^k)(x^l-c^l)
\label{eq:KijAxisymReflReducedQart}
\end{align}
where  $\bm b = b\,\bm{\hat z}, \bm c = c\,\bm{\hat z}$ and $a,b,c$ are arbitrary parameters.
It will be shown in Sec. \ref{sec:34} that $\mu=0$, leaving the first term $A^{(ij}B^{kl)}$ in Eq. \eqref{eq:KillingTensorRankFour} as the only possibility for an irreducible Killing tensor of rank-4. Note that the above expressions were obtained in two dimensions. In three dimensions, additional reducible terms constructed from combinations of the axial Killing vector $\varphi^i$ and other Killing tensors ($\varphi^j$ or $g^{kl}$) will be present, which vanish for meridional motion. These reducible terms can be dropped, by virtue of energy and angular momentum conservation. The  combination $A^{(ij}B^{kl)}$ is thus the only nontrivial possibility in both two and three dimensions.
 This completes step (i) of our prescribed procedure.

\subsection{First integrability condition: no-hair relation} \label{sec:nohairquartic}
\label{}

With the tensor $K^{ijkl}$ known, the next step consists of finding the class of  gravitational potentials $\Phi$ that satisfy the integrability condition  \eqref{eq:IntegrabilityCondQuart2}. 
A stationary axisymmetric vacuum potential that vanishes at infinity is characterized by the multipole expansion \eqref{eq:MultipoleExpansion}. Assuming equatorial plane reflection symmetry, the odd moments  $M_L$ $(L=1,3,...)$  vanish. Restricting attention to purely meridional motion, with  $K^{ijkl}$ and  $\Phi$ given by eqs. \eqref{eq:KillingTensorRankFour} and  \eqref{eq:MultipoleExpansion}, we find with straightforward algebra that the integrability condition  \eqref{eq:IntegrabilityCondQuart2} is satisfied if and only if the even multipole moments satisfy the two-step recursion relation
\begin{equation} \label{eq:nohairquarticrecursion1}
\kappa M_{L+4}=\left[\kappa (a^2+b^2)-\frac{\mu}{8}\right]M_{L+2}-\left(\kappa a^2b^2-\frac{\mu}{8}c^2 \right)M_L 
\end{equation}

If $\kappa=0$ and $\mu\neq0$, the above relation reduces to the ``no-hair'' relation  \eqref{eq:MomentsRecursionOne} and Eq. \eqref{eq:KillingTensorRankFour}
gives a reducible   rank-4 Killing tensor, constructed from the  rank-2 Killing tensor \eqref{eq:MultipoleExpansion} and the metric $g^{ij}$.

If $\kappa\neq0$ and $\mu=0$, the above relation simplifies to
\begin{equation} \label{eq:nohairquarticrecursion}
M_{L+4}=(a^2+b^2)M_{L+2}-a^2b^2 M_L 
\end{equation}
which generalizes the ``no-hair'' relation  \eqref{eq:MomentsRecursionOne}.
  One may obtain  all higher  moments recursively from the first two nonvanishing moments. This yields
\begin{equation} \label{eq:nohairquartic}
M_{L}=\frac{(a^2b^L-a^Lb^2)M_0+(a^L-b^L)M_2}{a^2-b^2} 
\end{equation}
for $L$ even (and $M_L=0$ for $L$ odd).
In the limit $b\rightarrow 0$, one recovers the no-hair relations  \eqref{eq:MomentsRecursionOne}
and \eqref{eq:MomentsMa} (except for  $L=0$). 
The limit $b\rightarrow a$ is slightly more subtle. 
Although the $b\rightarrow a$ limit of eq.
\eqref{eq:nohairquarticrecursion}
is satisfied by  \eqref{eq:MomentsRecursionOne} 
the converse is not necessarily true,  because Eq. 
\eqref{eq:nohairquarticrecursion}  does not fix the relation between the first two moments.
Applying Eq. 
\eqref{eq:nohairquarticrecursion}
recursively
after taking the limit  $b\rightarrow a$ (or taking the same limit of Eq. \eqref{eq:nohairquartic}
and
applying the l' H\^ospital rule)   yields
\begin{equation} \label{eq:nohairquarticaequalsb}
\lim_{b\rightarrow a}  M_{L}=a^L\left[M_0+\frac{L}{2}(a^{-2}M_2-M_0)\right] 
\end{equation}
If the mass and quadrupole moment are related by 
$M_2=a^2 M_0$ then one recovers 
Eq.  \eqref{eq:MomentsMa}, but this need not be the case and $M_2$
is generally considered independent of $M_0$.
If $a\neq b$, then the gravitational field depends on  four independent  parameters  ($M_0, M_2, a, b$) and the reparametrization
$m_a=(M_2-b^2 M_0)/(a^2-b^2)$, $m_b=(M_2-a^2 M_0)/(b^2-a^2)$ allows us to write Eq. \eqref{eq:nohairquartic} in the suggestive form
\begin{equation} \label{eq:nohairquartic4masses}
M_{L}=a^L m_{a} + b^L m_{b}  
\end{equation}
Then, summing the Legendre series  \eqref{eq:MultipoleExpansion}
yields the potential  
\begin{align} \label{eq:PhiQuadrapole}
\Phi=&-\frac{m_{a}/2}{\sqrt{x^2+y^2+(z-a)^2}}
-\frac{m_{a}/2}{\sqrt{x^2+y^2+(z+a)^2}}
\nonumber\\&-\frac{m_{b}/2}{\sqrt{x^2+y^2+(z-b)^2}}
-\frac{m_{b}/2}{\sqrt{x^2+y^2+(z+b)^2}}
\end{align}
created by four fixed point sources with mass $m_{a}/2$  at positions $\pm a \, \bm {\hat z}$ and  mass
 $m_{b}/2$  at positions $\pm b \, \bm {\hat z}$.

If $\kappa\neq0$ and $\mu\neq0$, one may introduce new parameters $\alpha,\beta$
such that $\kappa (a^2+b^2)-\frac{\mu}{8}=\kappa (\alpha^2+\beta^2)$ and $\kappa a^2b^2-\frac{\mu}{8}c^2=\kappa \alpha^2 \beta^2$. Eq. \eqref{eq:nohairquarticrecursion1} then takes the form
\begin{equation} \label{eq:nohairquarticrecursion3}
M_{L+4}=(\alpha^2+\beta^2)M_{L+2}-\alpha^2 \beta^2 M_L 
\end{equation}
which is analogous to Eq. \eqref{eq:nohairquarticrecursion}. One can then proceed as in the previous case and obtain the multipole moments
\begin{equation} \label{eq:nohairquartic4masses2}
M_{L}=\alpha^L m_{\alpha} + \beta^L m_{\beta}  
\end{equation}
of a potential analogous to  
\eqref{eq:PhiQuadrapole}, with $a,b$ replaced by $\alpha,\beta$.

Our solution procedure guarantees that Eq.   
 \eqref{eq:nohairquartic4masses} (or its analogue for nonzero $\mu$, Eq. \eqref{eq:nohairquartic4masses2}) gives \textit{ the unique vacuum potential with the symmetry $(\mathbb{R},+)\times \rm{SO(2)} \times \mathbb{Z}_2$ compatible with the integrability condition   \eqref{eq:IntegrabilityCondQuart2}.} This completes step (ii) of the prescribed procedure.

\subsection{Second integrability condition: nonexistence of quartic invariants} \label{sec:34}

With $\Phi$ obtained from  Eq. \eqref{eq:PhiQuadrapole} ($\kappa\neq0$ and $\mu=0$) or \eqref{eq:nohairquartic4masses2} ($\kappa\neq0$ and $\mu\neq0$), one may use  
eqs. \eqref{eq:Kyyinhom}-\eqref{eq:Kzyinhom} to solve the inhomogeneous Killing equation \eqref{eq:KillingEq4th2}. 

In the  case $\kappa\neq0$,  $\mu=0$, we find
\begin{align}
{K^{ij}} =& 2B{A^{ij}} + 2A{B^{ij}} 
\nonumber\\+& ({b^2} - {a^2})(\tilde A - \tilde B)(z\delta _y^i\delta _y^j - y\delta _y^{(i}\delta _z^{j)})+\nu \, D^{ij} \label{eq:KijInhomSolution2D}
\end{align}
where $\nu$ is a constant, $A$ is given by Eq. \eqref{eq:KDipole} with the replacement $M_0\rightarrow2m_a$, $\tilde A$ is given by 
\begin{equation}
\tilde A = \frac{{{{m_a}}(z+a)}}{{\sqrt {{y^2} + {{(z+a )}^2}} }} + \frac{{{{m_a}}(z - a)}}{{\sqrt {{y^2} + {{(z - a)}^2}} }}
\label{eq:AbarInhomSol2D}
\end{equation}
and $B$, $\tilde B$ are given by Eqs. \eqref{eq:KDipole}, \eqref{eq:AbarInhomSol2D}
 with the replacements $a\rightarrow b$ and $m_a \rightarrow m_b$.
The additive contribution $C^{ij}$ to Eq. \eqref{eq:KijInhomSolution2D} is a solution to the homogenous Killing equation  \eqref{eq:KillingEquationsQuad},
which is  polynomial of second order in the cartesian coordinates. Since the homogeneous solution must obey the symmetries of the problem, it must have  the form of  Eq. \eqref{eq:KijAxisymReflReduced}, with $a$ replaced by some other parameter $c$, that is
\begin{equation}  \label{eq:CijAxisymReflReduced}
{D^{ij}} 
=g^{mn} \varepsilon^i_{\phantom{i}km} \varepsilon^j_{\phantom{j}ln} 
(x^k+d^k)(x^l-d^l) 
\end{equation}
where  $\bm d=d \, \hat{\bm z}$.

In the case $\kappa\neq0$,  $\mu\neq0$, one may proceed in a similar way. However, upon substitution of the resulting expressions into 
$\eqref{eq:Kyzinhom}, \eqref{eq:Kzyinhom}$, we find that the two relations yield different  results, unless $\mu=0$. Thus, since $K^{ij}$ must be symmetric, we are  left with $\kappa\neq0$, $\mu=0$, and Eq. \eqref{eq:nohairquarticrecursion},  as the only viable possibility.
This completes step (iii) of our prescribed procedure.

 The next step is to solve the second integrability condition \eqref{eq:IntegrabilityCondQuart3} for a potential with moments  given by Eq. \eqref{eq:nohairquarticrecursion}.  However, substituting Eqs.
\eqref{eq:PhiQuadrapole} and  \eqref{eq:KijInhomSolution2D}
into the condition \eqref{eq:IntegrabilityCondQuart3}, we find that the 
latter \textit{cannot be satisfied} except in the limit $b \rightarrow a$, whence the quartic invariant is reducible to the quadratic invariant of the previous section.

We infer that  \textit{there exists no stationary, axisymmetric, equatorially symmetric  vacuum Newtonian potential that admits an  independent nontrivial invariant  quartic in the momenta}. That is, the only quartic invariants are trivial products of lower-order invariants.  

The proof of this nonexistence result used the fact that strong integrability requires existence of a constant of motion for all values of  energy $E$ and angular momentum $L_z$, including the case of purely meridional orbits with $L_z=0$.
Failure
to satisfy Eq.  \eqref{eq:IntegrabilityCondQuart2}  means that there exists no independent constant for purely meridional motion; therefore there exists no strong integral for three-dimensional motion. One could  alternatively use
the integrability condition  \eqref{eq:IntegrabilityCondQuart22} to derive the same nonexistence result directly in $\mathbb{E}^3$.

As mentioned earlier, this section was motivated by the  intuitive expectation that 
  4-centre potentials may admit rank-4 Killing tensors, reducible to a combination of rank-2 tensors in the appropriate limit. Remarkably, this expectation was  partially fulfilled, in the sense that the most general rank-4 tensor consistent with stationarity,  axisymmetry and equatorial  symmetry, given by Eq.  \eqref{eq:KillingTensorRankFour}, is precisely such a combination and that  its integrability conditions  \eqref{eq:IntegrabilityCondQuart2}  lead uniquely  to  4-centre potentials, given by Eq. \eqref{eq:PhiQuadrapole}. Nevertheless, since the remaining integrability conditions
\eqref{eq:IntegrabilityCondQuart3} are not satisfied, this tensor does not give rise to a quartic invariant.
Simple considerations based on Poisson brackets  show that integrability is a non-linear property, in the sense that a linear superposition of integrable Newtonian potentials need not be also integrable.
In view of this, the integrability of 2-centre potentials (and their relativistic
analogues) is exceptional, 
but the non-integrability of 4-centre potentials is not  surprising.

The nonexistence of a quartic invariant does not nevertheless preclude the possibility that the potential \eqref{eq:PhiQuadrapole} admits some other constant of motion with different dependence in the momenta.
 A superficial study of Poincar\'e maps  of three dimensional orbits in the potential  \eqref{eq:PhiQuadrapole} may show that most orbits appear to be regular. This could lead to the (false) impression that the system is integrable.  However, a thorough scan of initial conditions reveals the existence of \textit{Birkhoff chains} and in some cases  \textit{ergodic motion} surrounds the main island of stability on the Poincar\'e maps, confirming the non-integrability of 4-centre potentials. 
In a relativistic context, a similar behaviour has been observed for orbits in certain stationary axisymmetric spacetimes 
\citep{Apostolatos2009,Lukes2010,CONTOPOULOS2011}.

\section{Summary and Conclusions} 
\label{sec:summary} 
The aim of this work was to use a direct approach in order to study polynomial constants of motion in  stationary axisymmetric gravitational fields in a  Newtonian context and to gain insight into  analogous  problems in relativistic gravity. The results established via this direct search   include:  
\begin{enumerate}
\item
the uniqueness of   the constant \eqref{eq:KDipole}-\eqref{eq:quadinvariantdipole} of the Euler problem: Eq.   \eqref{eq:PhiDipole} gives the only 
stationary, axisymmetric, equatorially symmetric Newtonian vacuum potential
admitting a Killing-St\"ackel tensor of rank two;

\item
the uniqueness of the constant \eqref{eq:quadinvariantdipole}-\eqref{eq:ADipoleLamda} of the Lagrange problem: Eq. \eqref{eq:PhiDipoleLamda} gives the only 
stationary, axisymmetric, equatorially symmetric Newtonian potential
admitting a Killing-St\"ackel tensor of rank two;

\item
the relation  of the quadratic constant in the Lagrange  problem to that of  \cite{Lynden-Bell2003}, and the analogy with  the Carter constant in a Kerr-de Sitter spacetime;

\item
the  integrability conditions \eqref{eq:IntegrabilityCondQuart2},
\eqref{eq:IntegrabilityCondQuart22}
which do not appear to have been  implemented previously in  direct searches for quartic invariants  (c.f. \citealt{Hietarinta1987}). 

\item
the non-existence of  stationary, axisymmetric, equatorially symmetric, vacuum Newtonian potentials admitting a  Killing-St\"ackel tensor of rank four.

\end{enumerate}
Note that the assumptions of vacuum and equatorial symmetry may be relaxed without loss of integrability in the two-center problem   \citep{Lynden-Bell2003}. Note also that the integrability condition
\eqref{eq:IntegrabilityCondQuart2} and its generalization to arbitrary dimension, Eq.
\eqref{eq:IntegrabilityCondQuart22},
can be  quite useful for other  direct searches of quartic invariants in physical systems.

Electromagnetic or gravitational-wave observations of  orbital motion around a massive black hole
 at the galactic centre can probe its  gravitational field and test the validity of the Kerr solution and the no-hair theorem. To this end, several authors have explored the possibility of  spacetime mapping by modelling the central object as a bumpy black hole (cf. the review by \citealt{Johannsen2011a} and references therein). To this end, 
\cite{Gair2008} and \cite{Brink2008,Brink2010,Brink2010a} suggested the possibility of using integrable stationary axisymmetric spacetimes of  the Manko-Novikov type. However, a conjecture on existence of a quartic constant of motion \citep{Brink2011} was later disproven  by \citep{Kruglikov2011} and \citep{Lukes2012}
for the case of the Zipoy-Voorhees metric. \cite{Mirshekari2010} provide a non-existence proof for the Bach-Weyl spacetime. Although  non-existence results have been obtained for particular stationary axisymmetric spacetimes, no  such result has been established for stationary axisymmetric spacetimes with arbitrary multipole moments.

The present work provides a non-existence proof of quartic invariants for  generic stationary axisymmetric vacuum gravitational fields in the Newtonian regime. Extending the present  analysis to the relativistic regime is a nontrivial task. However, the methods employed here  provide useful intuition for  treating the analogous problem in relativistic gravity or post-Newtonian 
approximations to it. In particular, if a stationary axisymmetric system is nonintegrable in the Newtonian limit, it is unlikely to be integrable in the relativistic regime (although the converse is not true). This is consistent with  \cite{Kruglikov2011} and \cite{Lukes2012} and provides evidence in favour of the conjecture that a stationary, axisymmetric,  equatorially symmetric, vacuum (modulo a cosmological constant) spacetime is integrable if and only if it belongs to the Kerr (or Kerr-de Sitter) class. Therefore, attempts to 
seek stationary axisymmetric and equatorially symmetric vacuum solutions in general relativity  (other than Kerr) that admit irreducible polynomial  invariants are unlikely to yield positive results.

 This can lead to further insights on how to parametrize the  departure of a  spacetime geometry from that of a Kerr spacetime and the relation  of this departure to integrabilty or ergodocity. In particular, if the nonexistence conjecture is true, at least two conclusions can be drawn. First,   ergodic geodesics   exist if and only if the spacetime geometry deviates from Kerr. This is in agreement  with numerical evidence  in \cite{Apostolatos2009,Lukes2010,Lukes2010a,CONTOPOULOS2011}, although a general proof is still lacking.  Second, approaches towards modelling bumpy black holes as integrable, such as those attempted by \cite{Gair2008,Brink2008,Brink2010,Brink2010a} 
are less likely to be successful than  approaches that do not require integrability, such as those based on canonical perturbation theory  \citep{Vigeland2010a,Vigeland2010}.

As mentioned earlier, the uniqueness and non-existence proofs for Newtonian gravity   are strongly suggestive of similar behaviour in General Relativity.  A rotating black hole's mass moments are identical to those of  two Newtonian  centers fixed at  imaginary distance from each other. Thus, the Euler (and Lagrange) problems  can   qualitatively  capture many features (such as  integrability  of motion, cosmological constant, resonant frequencies, separability of wave equations and other non-gravitomagnetic phenomena) of their relativistic counter-parts. Newtonian  systems
can therefore serve as simple toy-models whose  study as surrogates of Kerr and non-Kerr spacetimes offers  valuable insights  and opens interesting questions. For example, in Newtonian gravity, the quadratic constant is known to extend beyond  equatorial plane symmetry   
(\citealt*{Aksenov1963}; \citealt{Lukyanov2005a,Lynden-Bell2003}).
Since the Kerr family is equatorially symmetric, this raises the question of whether other stationary  axisymmetric solutions to the vacuum Einstein equations exist  that  still admit a quadratic constant  of motion but  are not equatorially symmetric.

\section*{Acknowledgments}

The author  thanks  T. Apostolatos, J. Brink, B. Carter, \'E.~\'E. Flanagan, J.~L. Friedman, K. Glampedakis
and G. Pappas and the anonymous referees for  fruitful suggestions and comments. The author also thanks N. K. Johnson-McDaniel for  corrections to the manuscript and  for  generalizing Eq. \eqref{eq:IntegrabilityCondQuart2} to  Eq. \eqref{eq:IntegrabilityCondQuart22} and  G. Lukes-Gerakopoulos for  useful comments and for providing Poincar\'e maps of the four-center potential. This work was supported  in part  by the Greek State Scholarships Foundation, by
NSF Grant PHY1001515  and  by DFG grant SFB/Transregio 7
``Gravitational Wave Astronomy''.

\setlength{\bibhang}{2.0em}
\setlength\labelwidth{0.0em}
\bibliographystyle{mn2e}
\bibliography{library}

\begin{thebibliography}{78}
\expandafter\ifx\csname natexlab\endcsname\relax\def\natexlab#1{#1}\fi

\bibitem[{Aksenov {et~al}\mbox{.}(1963)Aksenov, Grebenikov, \&
  Demin}]{Aksenov1963}
Aksenov E.~P., Grebenikov E.~A., Demin V.~G., 1963, Soviet Astronomy, 7, 491

\bibitem[{Amaro {et~al}\mbox{.}(2012{\natexlab{a}})Amaro, Aoudia, Babak,
  Binetruy, Berti, Bohe, Caprini, Colpi, Cornish, Danzmann, Dufaux, Gair,
  Jennrich, Jetzer, Klein, Lang, Lobo, Littenberg, McWilliams, Nelemans,
  Petiteau, Porter, Schutz, Sesana, Stebbins, Sumner, Vallisneri, Vitale,
  Volonteri, \& Ward}]{Amaro2012a}
Amaro P. {et~al.}, 2012{\natexlab{a}}, arXiv:1202.0839, 20

\bibitem[{Amaro {et~al}\mbox{.}(2012{\natexlab{b}})Amaro, Aoudia, Babak,
  Bin\'{e}truy, Berti, Boh\'{e}, Caprini, Colpi, Cornish, Danzmann, Dufaux,
  Gair, Jennrich, Jetzer, Klein, Lang, Lobo, Littenberg, McWilliams, Nelemans,
  Petiteau, Porter, Schutz, Sesana, Stebbins, Sumner, Vallisneri, Vitale,
  Volonteri, \& Ward}]{Amaro2012}
Amaro P. {et~al.}, 2012{\natexlab{b}}, arXiv:1201.3621

\bibitem[{Apostolatos {et~al}\mbox{.}(2009)Apostolatos, Lukes-Gerakopoulos, \&
  Contopoulos}]{Apostolatos2009}
Apostolatos T., Lukes-Gerakopoulos G., Contopoulos G., 2009, Physical Review
  Letters, 103, 111101

\bibitem[{Boccaletti \& Pucacco(2003)}]{Boccaletti2003}
Boccaletti D., Pucacco G., 2003, {Theory of Orbits: Volume 1: Integrable
  Systems and Non-perturbative Methods}. Springer, p. 406

\bibitem[{Brink(2008{\natexlab{a}})}]{Brink2008}
Brink J., 2008{\natexlab{a}}, Physical Review D, 78, 102001

\bibitem[{Brink(2008{\natexlab{b}})}]{Brink2008a}
Brink J., 2008{\natexlab{b}}, Physical Review D, 78, 102002

\bibitem[{Brink(2010{\natexlab{a}})}]{Brink2010}
Brink J., 2010{\natexlab{a}}, Physical Review D, 81, 022001

\bibitem[{Brink(2010{\natexlab{b}})}]{Brink2010a}
Brink J., 2010{\natexlab{b}}, Physical Review D, 81, 022002

\bibitem[{Brink(2011)}]{Brink2011}
Brink J., 2011, Physical Review D, 84, 104015

\bibitem[{Carter(1968)}]{Carter1968}
Carter B., 1968, Communications in Mathematical Physics (1965-1997), 10, 280

\bibitem[{Carter(1977)}]{Carter1977}
Carter B., 1977, Physical Review D, 16, 3395

\bibitem[{Carter(2009{\natexlab{a}})}]{CarterPC}
Carter B., 2009{\natexlab{a}}, Personal communication

\bibitem[{Carter(2009{\natexlab{b}})}]{Carter2009}
Carter B., 2009{\natexlab{b}}, General Relativity and Gravitation, 41, 2873

\bibitem[{Carter(2010)}]{Carter2010}
Carter B., 2010, General Relativity and Gravitation, 42, 653

\bibitem[{Collins \& Hughes(2004)}]{Collins2004}
Collins N., Hughes S., 2004, Physical Review D, 69, 124022

\bibitem[{Contopoulos {et~al}\mbox{.}(2011)Contopoulos, Lukes-Gerakopoulos, \&
  Apostolatos}]{CONTOPOULOS2011}
Contopoulos G., Lukes-Gerakopoulos G., Apostolatos T.~A., 2011, International
  Journal of Bifurcation and Chaos, 21, 2261

\bibitem[{de~Zeeuw(1985{\natexlab{a}})}]{DeZeeuw1985c}
de~Zeeuw T., 1985{\natexlab{a}}, Monthly Notices of the Royal Astronomical
  Society, 216, 599

\bibitem[{de~Zeeuw(1985{\natexlab{b}})}]{DeZeeuw1985b}
de~Zeeuw T., 1985{\natexlab{b}}, Monthly Notices of the Royal Astronomical
  Society, 216, 273

\bibitem[{de~Zeeuw(1985{\natexlab{c}})}]{DeZeeuw1985a}
de~Zeeuw T., 1985{\natexlab{c}}, Monthly Notices of the Royal Astronomical
  Society, 215, 731

\bibitem[{Dubovsky {et~al}\mbox{.}(2007)Dubovsky, Tinyakov, \&
  Zaldarriaga}]{Dubovsky2007}
Dubovsky S., Tinyakov P., Zaldarriaga M., 2007, Journal of High Energy Physics,
  2007, 083

\bibitem[{Eddington(1915)}]{Eddington1915}
Eddington A.~S., 1915, Monthly Notices of the Royal Astronomical Society, 76,
  37

\bibitem[{Euler(1760)}]{Euler1760}
Euler L., 1760, Mem. Berlin, 228

\bibitem[{Euler(1764)}]{Euler1764}
Euler L., 1764, Novi Commet. Acad. Scient. Imperial. Petropolit., 10, 207

\bibitem[{Flanagan \& Hinderer(2007)}]{Flanagan2007}
Flanagan E., Hinderer T., 2007, Physical Review D, 75, 124007

\bibitem[{Gair {et~al}\mbox{.}(2008)Gair, Li, \& Mandel}]{Gair2008}
Gair J., Li C., Mandel I., 2008, Physical Review D, 77

\bibitem[{Glampedakis \& Babak(2006)}]{Glampedakis2006}
Glampedakis K., Babak S., 2006, Classical and Quantum Gravity, 23, 4167

\bibitem[{Hansen(1974)}]{Hansen1974}
Hansen R.~O., 1974, Journal of Mathematical Physics, 15, 46

\bibitem[{Hietarinta(1987)}]{Hietarinta1987}
Hietarinta J., 1987, Physics Reports, 147, 87

\bibitem[{Horwood(2008)}]{HORWOOD2008}
Horwood J., 2008, Journal of Geometry and Physics, 58, 487

\bibitem[{Hughes(2006)}]{Hughes2006}
Hughes S.~A., 2006in , AIP, pp. 233--240

\bibitem[{Ioannou \& Apostolatos(2004)}]{Ioannou2004}
Ioannou P.~J., Apostolatos T.~A., 2004, {Elements of Theoretical Mechanics (in
  Greek)}, 1st edn. Leader Books, Athens, p. 421

\bibitem[{Israel(1970)}]{Israel1970}
Israel W., 1970, Physical Review D, 2, 641

\bibitem[{Jacobi(2009)}]{Jacobi2009a}
Jacobi C. G.~J., 2009, {Jacobi's Lectures on Dynamics}, 2nd edn., Clebsch A.,
  ed. Hindustan Book Agency, New Delhi, p. 339

\bibitem[{Jennrich {et~al}\mbox{.}(2011)Jennrich, Binetruy, Colpi, Danzmann,
  Jetzer, Lobo, Nelemans, Schutz, Stebbins, Sumner, Vitale, \&
  Ward}]{OliverJennrich}
Jennrich O. {et~al.}, 2011, {NGO assessment study report (Yellow Book)}. Tech.
  rep.

\bibitem[{Johannsen(2012)}]{Johannsen2011a}
Johannsen T., 2012, Advances in Astronomy, 2012

\bibitem[{Johannsen \& Psaltis(2010{\natexlab{a}})}]{Johannsen2010b}
Johannsen T., Psaltis D., 2010{\natexlab{a}}, The Astrophysical Journal, 716,
  187

\bibitem[{Johannsen \& Psaltis(2010{\natexlab{b}})}]{Johannsen2010a}
Johannsen T., Psaltis D., 2010{\natexlab{b}}

\bibitem[{Johannsen \& Psaltis(2011)}]{Johannsen2011b}
Johannsen T., Psaltis D., 2011, Physical Review D, 83, 124015

\bibitem[{Johannsen \& Psaltis(2013)}]{Johannsen2013}
Johannsen T., Psaltis D., 2013, The Astrophysical Journal, 773, 57

\bibitem[{Kalnins(2012)}]{Kalnins2012}
Kalnins E.~G., 2012, Symmetry, Integrability and Geometry: Methods and
  Applications, 8, 034

\bibitem[{Kalnins {et~al}\mbox{.}(2010)Kalnins, Kress, \& Miller}]{Kalnins2010}
Kalnins E.~G., Kress J.~M., Miller W., 2010, Journal of Physics A: Mathematical
  and Theoretical, 43, 265205

\bibitem[{Kalnins {et~al}\mbox{.}(2009)Kalnins, Kress, \& {Miller
  Jr}}]{Kalnins2009}
Kalnins E.~G., Kress J.~M., {Miller Jr} W., 2009

\bibitem[{Keres(1967)}]{Keres1967}
Keres H., 1967, Soviet Journal of Experimental and Theoretical Physics, 25, 504

\bibitem[{Kerr \& Schild(1965)}]{Kerr}
Kerr R.~P., Schild A., 1965, Proc. Symp. Appl. Math., 17

\bibitem[{Kraniotis(2004)}]{Kraniotis2004}
Kraniotis G.~V., 2004, Classical and Quantum Gravity, 21, 4743

\bibitem[{Kraniotis(2005)}]{Kraniotis2005}
Kraniotis G.~V., 2005, Classical and Quantum Gravity, 22, 4391

\bibitem[{Kraniotis(2011)}]{Kraniotis2011}
Kraniotis G.~V., 2011, Classical and Quantum Gravity, 28, 085021

\bibitem[{Kruglikov \& Matveev(2011)}]{Kruglikov2011}
Kruglikov B.~S., Matveev V.~S., 2011, arXiv:1111.4690

\bibitem[{Kuzmin(1956)}]{Kuzmin1956}
Kuzmin G.~G., 1956, Astr. Zh., 33

\bibitem[{Lagrange(1766)}]{Lagrange1766}
Lagrange J.-L., 1766, Oeuvres de Lagrange, 2, 67

\bibitem[{Lukes {et~al}\mbox{.}(2010)Lukes, Apostolatos, \&
  Contopoulos}]{Lukes2010a}
Lukes G., Apostolatos T.~A., Contopoulos G., 2010, Physical Review D, 81, 25

\bibitem[{Lukes-Gerakopoulos(2012)}]{Lukes2012}
Lukes-Gerakopoulos G., 2012, Physical Review D, 86, 044013

\bibitem[{Lukes-Gerakopoulos {et~al}\mbox{.}(2010)Lukes-Gerakopoulos,
  Apostolatos, \& Contopoulos}]{Lukes2010}
Lukes-Gerakopoulos G., Apostolatos T.~A., Contopoulos G., 2010, Physical Review
  D, 81, 124005

\bibitem[{Lukyanov {et~al}\mbox{.}(2005)Lukyanov, Emeljanov, \&
  Shirmin}]{Lukyanov2005a}
Lukyanov L.~G., Emeljanov N.~V., Shirmin G.~I., 2005, Cosmic Research, 43, 186

\bibitem[{Lynden-Bell(1962)}]{Lynden-Bell1962}
Lynden-Bell D., 1962, Monthly Notices of the Royal Astronomical Society, 124,
  95

\bibitem[{Lynden-Bell(2003)}]{Lynden-Bell2003}
Lynden-Bell D., 2003, Monthly Notices of the Royal Astronomical Society, 338,
  208

\bibitem[{Mirshekari \& Will(2010)}]{Mirshekari2010}
Mirshekari S., Will C.~M., 2010, Classical and Quantum Gravity, 27, 235021

\bibitem[{Misner {et~al}\mbox{.}(1973)Misner, Thorne, \& Wheeler}]{Misner1973}
Misner C.~W., Thorne K.~S., Wheeler J.~A., 1973, {Gravitation}. W. H. Freeman,
  p. 1215

\bibitem[{\'{O}'Math\'{u}na(2008)}]{OMathuna2008}
\'{O}'Math\'{u}na D., 2008, {Integrable Systems in Celestial Mechanics}.
  Birkh\"{a}user Boston, p. 244

\bibitem[{Padmanabhan(2010)}]{Padmanabhan2010}
Padmanabhan T., 2010, {Gravitation: Foundations and Frontiers}. Cambridge
  University Press, p. 728

\bibitem[{Psaltis \& Johannsen(2009)}]{Psaltis2009a}
Psaltis D., Johannsen T., 2009, 4

\bibitem[{Psaltis \& Johannsen(2011)}]{Psaltis2011}
Psaltis D., Johannsen T., 2011, Journal of Physics: Conference Series, 283,
  012030

\bibitem[{Psaltis \& Johannsen(2012)}]{Psaltis2012}
Psaltis D., Johannsen T., 2012, The Astrophysical Journal, 745, 1

\bibitem[{Ryan(1995)}]{Ryan1995}
Ryan F., 1995, Physical Review D, 52, 5707

\bibitem[{Sotiriou \& Apostolatos(2005)}]{Sotiriou2004}
Sotiriou T.~P., Apostolatos T.~A., 2005, Phys. Rev. D, 71, 044005

\bibitem[{St\"{a}ckel(1890)}]{Stackel1890}
St\"{a}ckel P., 1890, Math. Ann., 35

\bibitem[{Trahanas(2004)}]{Trahanas2004}
Trahanas S., 2004, {Partial Differential Equations (in Greek)}. Crete
  University Press, Heraklion, Crete

\bibitem[{Vigeland(2010)}]{Vigeland2010a}
Vigeland S., 2010, Physical Review D, 82, 104041

\bibitem[{Vigeland {et~al}\mbox{.}(2011)Vigeland, Yunes, \&
  Stein}]{Vigeland2011}
Vigeland S., Yunes N., Stein L., 2011, Physical Review D, 83, 104027

\bibitem[{Vigeland \& Hughes(2010)}]{Vigeland2010}
Vigeland S.~J., Hughes S.~A., 2010, Physical Review D, 81, 024030

\bibitem[{Vinti(1960)}]{Vinti1960}
Vinti J.~P., 1960, The Astronomical Journal, 65, 353

\bibitem[{Vinti(1963)}]{Vinti1963}
Vinti J.~P., 1963, in The Use of Artificial Satellites for Geodesy, Veis G.,
  ed., Interscience Publishers, Amsterdam, p.~12

\bibitem[{Vinti(1969)}]{Vinti1969}
Vinti J.~P., 1969, The Astronomical Journal, 74, 25

\bibitem[{Vinti(1971)}]{Vinti1971}
Vinti J.~P., 1971, Celestial Mechanics, 4, 348

\bibitem[{Vinti {et~al}\mbox{.}(1998)Vinti, Der, \& Bonavito}]{Vinti1998}
Vinti J.~P., Der G.~J., Bonavito N.~L., 1998, {Orbital and celestial
  mechanics}. American Institute of Aeronautics and Astronautics, p. 409

\bibitem[{Whittaker(1989)}]{Whittaker1989}
Whittaker E.~T., 1989, {A Treatise on the Analytical Dynamics of Particles and
  Rigid Bodies}. Cambridge University Press, p. 480

\bibitem[{Will(2009)}]{Will2009}
Will C., 2009, Physical Review Letters, 102

\end{thebibliography}


\begin{thebibliography}{99}
\bibitem[\protect\citeauthoryear{Baird}{1981}]{b1} Baird S.R., 1981,
ApJ, 245, 208
\bibitem[\protect\citeauthoryear{Beichman et al.}{1985a}]{b2} Beichman
C.A., Neugebauer G., Habing H.J., Clegg P.E., Chester T.J., 1985a,
{\it IRAS\/} Point Source Catalog. Jet Propulsion Laboratory,
Pasadena
\bibitem[\protect\citeauthoryear{Beichman et al.}{1985b}]{b3} Beichman
C.A., Neugebauer G., Habing H.J., Clegg P.E., Chester T.J., 1985b,
{\it IRAS\/} Explanatory Supplement. Jet Propulsion Laboratory,
Pasadena
\bibitem[\protect\citeauthoryear{Dawson}{1979}]{b4} Dawson D.W., 1979,
ApJS, 41, 97
\bibitem[\protect\citeauthoryear{Gerhz}{1972}]{b5} Gerhz R.D., 1972, ApJ,
178, 715
\bibitem[\protect\citeauthoryear{Gerhz \& Ney}{1972}]{b6} Gerhz R.D., Ney
E.P., 1972, PASP, 84, 768
\bibitem[\protect\citeauthoryear{Gerhz \& Woolf}{1970}]{b7} Gerhz R.D., Woolf N.J.,
1970, ApJ, 161, L213
\bibitem[\protect\citeauthoryear{Gilman}{1972}]{b8} Gilman R.C., 1972, ApJ, 178, 423
\bibitem[\protect\citeauthoryear{Goldsmith et al.}{1987}]{b9} Goldsmith M.J., Evans A.,
Albinson J.S., Bode M.F., 1987, MNRAS, 227, 143
\bibitem[\protect\citeauthoryear{Hacking et al.}{1985}]{b10} Hacking P. et al., 1985,
PASP, 97, 616
\bibitem[\protect\citeauthoryear{Harvey, Thronson \& Gatley}{Harvey et al.}{1979}]{b11}
Harvey P.M., Thronson H.A., Gatley I., 1979, ApJ, 231, 115
\bibitem[\protect\citeauthoryear{Jura}{1986}]{b12} Jura M., 1986, ApJ, 309, 732
\bibitem[\protect\citeauthoryear{Kukarkin et al.}{1969}]{b13} Kukarkin B.V. et al.,
1969, General Catalogue of Variable Stars. Moscow
\bibitem[\protect\citeauthoryear{Lloyd Evans}{1974}]{b14} Lloyd Evans T., 1974, MNRAS,
167, 17{\sc p}
\bibitem[\protect\citeauthoryear{Lloyd Evans}{1985}]{b15} Lloyd Evans T., 1985, MNRAS,
217, 493
\bibitem[\protect\citeauthoryear{Low et al.}{1984}]{b16} Low F.J. et al., 1984, ApJ,
278, L19
\bibitem[\protect\citeauthoryear{McLaughlin}{1932}]{b17} McLaughlin D.B., 1932, Publ. Univ.
Obs. Mich., 4, 135
\bibitem[\protect\citeauthoryear{O'Connell}{1961}]{b18} O'Connell J.K., 1961, Specola
Vaticana Ric. Astron., 6, 341
\bibitem[\protect\citeauthoryear{Olnon \& Raimond}{1986}]{b19} Olnon F.M., Raimond E., 1986,
A\&AS, 65, 607
\bibitem[\protect\citeauthoryear{Preston et al.}{1963}]{b20} Preston G.W., Krzeminski W., Smak J.,
Williams J.A., 1963, ApJ, 137, 401
\bibitem[\protect\citeauthoryear{Rowan-Robinson \& Harris}{1983a}]{b21} Rowan-Robinson M., Harris
S., 1983a, MNRAS, 202, 767
\bibitem[\protect\citeauthoryear{Rowan-Robinson \& Harris}{1983b}]{b22} Rowan-Robinson M., Harris
S., 1983b, MNRAS, 202, 797
\bibitem[\protect\citeauthoryear{van der Veen \& Habing}{1988}]{b23} van der Veen W.E.C.J., Habing
H.J., 1988, A\&A, 194, 125
\bibitem[\protect\citeauthoryear{Willems \& de Jong}{1988}]{b24} Willems F.J., de Jong T., 1988,
A\&A, 196, 173
\bibitem[\protect\citeauthoryear{Zuckerman \& Dyck}{1986}]{b25} Zuckerman B., Dyck H.M., 1986, ApJ,
311, 345
\end{thebibliography}

\appendix

\section{integrability conditioNs and Inhomogeneous Killing equations }
\label{ch:appendixInhomKilling}

We consider the inhomogeneous Killing equation
\begin{equation}   \label{eq:inhomKillingR2}
{\partial _{(k}}{K_{ij)}} =f_{ijk}
\end{equation}
where $f^{ijk}=4{K^{ijkl}}{\partial _l}\Phi$. In $\mathbb{E}^2$, the cartesian components of the above equation  read
\begin{align}
{\partial _{y}}{K_{yy}} &=f_{yyy}  \label{eq:ihomKillingR21}   \\
{\partial _{z}}{K_{yy}}+{2\partial _{y}}{K_{yz}} &=3f_{yyz}  \label{eq:ihomKillingR22} \\
{\partial _{y}}{K_{zz}}+{2\partial _{z}}{K_{zy}} &=3f_{yzz}  \label{eq:ihomKillingR23} \\
{\partial _{z}}{K_{zz}} &=f_{zzz}  \label{eq:ihomKillingR24}
\end{align}
The above system is overdetermined and may be solved as follows: Integrating Eq. \eqref{eq:ihomKillingR21}
with respect to $y$ and Eq. \eqref{eq:ihomKillingR24} with respect to $z$ yields
\begin{align}   
{K_{yy}} = Z(z) + \int {dy{f_{yyy}}} \label{eq:Kyyinhom}  \\
{K_{zz}} = Y(y) + \int {dz{f_{zzz}}} \label{eq:Kzzinhom}
\end{align}
where $Z(z)$ and $Y(y)$ are  scalar functions of their arguments.
Then, integrating Eq. \eqref{eq:ihomKillingR22} with respect to $y$ and
Eq. \eqref{eq:ihomKillingR23} with respect to $z$ yields
\begin{align}   
{K_{yz}} = \zeta (z) + \frac{1}{2}\int {dy(3{f_{yyz}} - {\partial _z}{K_{yy}})} \label{eq:Kyzinhom}  \\
{K_{zy}} = \psi (y) + \frac{1}{2}\int {dz(3{f_{yzz}} - {\partial _y}{K_{zz}})} \label{eq:Kzyinhom}
\end{align}
where $\zeta(z)$ and $\psi(y)$ are   scalar functions of their arguments. Eqs.   \eqref{eq:Kyyinhom}-\eqref{eq:Kzyinhom} provide the solution to the inhomogeneous Killing equation   \eqref{eq:inhomKillingR2}.

Because $K_{ij}$ is a symmetric tensor, the above two expressions must be equal.  Acting with 
$\partial_{yyzz}=\frac{\partial^4}{\partial y^2 \partial z^2}$ on Eqs. \eqref{eq:Kyzinhom},
\eqref{eq:Kzyinhom},
demanding that the two expressions  be equal and using Eqs. 
\eqref{eq:Kyyinhom}, \eqref{eq:Kzzinhom},
yields
the integrability condition
\begin{equation}   \label{eq:inhomIntegrabilityCondition}
\partial_{zzz}f_{yyy}-3\partial_{zzy}f_{yyz}={\partial_{yyy}f_{zzz}} - 3\partial_{yyz}{f_{yzz}}
\end{equation}
which  must necessarily be satisfied by $f_{ijk}$ in order for 
Eq. \eqref{eq:inhomKillingR2} to have a solution.
Note that  the unknown functions 
$Z,Y,\zeta,\psi$ do not  appear in the above condition. In the homogeneous case, these functions can be easily shown to be quadratic polynomials in their arguments by setting  $f_{ijk}=0$, demanding that  expressions \eqref{eq:Kyzinhom} and
\eqref{eq:Kzyinhom} be equal and separating variables.






The integrability condition \eqref{eq:inhomIntegrabilityCondition} may be generalized to  $\mathbb{E}^n$  as follows. Eq. \eqref{eq:inhomKillingR2}
can be expanded out as
\begin{equation}   \label{eq:inhomKillingRn}
{\partial _{i}}{K_{jk}} +{\partial _{j}}{K_{ik}} +{\partial _{k}}{K_{ij}} =3f_{ijk}
\end{equation}
The first term vanishes  if we apply $\partial_l$ and antisymmetrize over $l$ and $i$. The second term vanishes if we subsequently apply  $\partial_m$ and antisymmetrize over $m$ and $j$. Finally, the third term vanishes  if we apply $\partial_n$ and antisymmetrize over $n$ and $k$. This yields the integrability condition
\begin{align}   \label{eq:inhomIntegrabilityConditionRn}
&\partial_{nml}f_{ijk}-\partial_{nmi}f_{ljk}-\partial_{njl}f_{imk}+\partial_{nji}f_{lmk}
\nonumber\\-&\partial_{kml}f_{ijn}+\partial_{kmi}f_{ljn}+\partial_{kjl}f_{imn}-\partial_{kji}f_{lmn}=0
\end{align}
which generalizes Eq.  \eqref{eq:inhomIntegrabilityCondition} to arbitrary dimension.

\section{ Killing-St\"ackel tensors OF Rank-four with 
reflection symmetry  in two dimensions}
\label{ch:appendixKillingRank4in2D}
As explained in Sec. \ref{rank4KStensors}, we are interested in solving Eq. \eqref{eq:KillingEq4th1} in $\mathbb{E}^2$. The most general solution,  according to Eq. \eqref{eq:KillingGeneralSolution} can be written as
\begin{align} \label{eq:Killing2DRank4Solution}
&\bm K =  
\mathcal{A}^{ijkl}{\bm X_i \!\otimes\! \bm X_j \!\otimes\!  \bm X_k \!\otimes\! \bm X_l}+ \mathcal{B}^{ijkl}{\bm X_i \!\otimes\! \bm X_j \!\otimes\!  \bm X_k \!\otimes\! \bm R_l}
\nonumber\\
&+\mathcal{C}^{ijkl}{\bm X_i \!\otimes\! \bm X_j \!\otimes\!  \bm R_k \!\otimes\! \bm R_l}+ \mathcal{D}^{ijkl}{\bm X_i \!\otimes\! \bm R_j \!\otimes\!  \bm R_k \!\otimes\! \bm R_l}
\nonumber\\
&+\mathcal{E}^{ijkl}{\bm R_i \!\otimes\! \bm R_j \!\otimes\!  \bm R_k \!\otimes\! 
\bm R_l} 
\end{align}
where the generators of translations and rotations are given by Eq. \eqref{eq:XiRi} and the constant coefficients possess the symmetries
$\mathcal{A}^{ijkl}=\mathcal{A}^{(ijkl)}$, 
$\mathcal{B}^{ijkl}=\mathcal{B}^{(ijk)l}$,
$\mathcal{C}^{ijkl}=\mathcal{C}^{(ij)(kl)}$,
$\mathcal{D}^{ijkl}=\mathcal{D}^{i(jkl)}$ and
$\mathcal{E}^{ijkl}=\mathcal{E}^{(ijkl)}$. Here, we consider  a meridional ($yz$) slice of the respective three-dimensional system. Indices are thus summed over values such that only $\bm e_y, \bm e_z$ and their linear combination $\bm R_x=y \,\bm e_z-z\, \bm e_y$ appear in $\bm K$; all other components vanish in the meridional plane. 
The number of independent coefficients is  reduced upon imposing symmetries.

First, we impose $\mathbb{Z}_2$ reflection symmetry about the equatorial ($xy$) plane by requiring that 
$\bm K$ be invariant under the replacement 
$\{z,\bm e_z \}\rightarrow \{-z,-\bm e_z \}$.
(This amounts to invariance of $K^{ijkl}p_i p_jp_kp_l$  under $\{z,p_z\}\rightarrow \{-z,-p_z\}$.) This yields the constraints
\begin{subequations} \label{eq:Z2constraints2D0}
\begin{align}  
\mathcal{A}^{yyyz}&=\mathcal{A}^{yzzz}=0 \label{eq:Z2constraints2D1}\\
\mathcal{B}^{yyzx}&=\mathcal{B}^{zzzx}=0 \label{eq:Z2constraints2D2}\\
\mathcal{C}^{yzxx}&=0 \label{eq:Z2constraints2D3} \\
\mathcal{D}^{zxxx}&=0  \label{eq:Z2constraints2D4}
\end{align}
\end{subequations}
Second,  recall that the  respective three-dimensional system is axisymmetric about the $z$-axis. Considering  a meridional ($yz$) slice of this system and making a rotation of $2 \pi$  radians about the $z$-axis amounts
to a reflection about the $xz$ plane. Thus, the respective two-dimensional system inherits a  
   $\mathbb{Z}_2$ reflection symmetry about the $xz$ plane. Imposing invariance with respect to the replacement 
$\{y,\bm e_y \}\rightarrow \{-y,-\bm e_y \}$
 yields the constraints
\begin{subequations} \label{eq:Z3constraints2D0}
\begin{align}  
\mathcal{A}^{yyyz}&=\mathcal{A}^{yzzz}=0 \label{eq:Z3constraints2D1}\\
\mathcal{B}^{yyyx}&=\mathcal{B}^{yzzx}=0 \label{eq:Z3constraints2D2}\\
\mathcal{C}^{yzxx}&=0 \label{eq:Z3constraints2D3} \\
\mathcal{D}^{yxxx}&=0  \label{eq:Z3constraints2D4}
\end{align}
\end{subequations}
Imposing both sets of constraints \eqref{eq:Z2constraints2D0},\eqref{eq:Z3constraints2D0}
to the tensor \eqref{eq:Killing2DRank4Solution} yields the solution
\begin{align} \label{eq:Killing2DRank4SolutionSimple}
&\bm K =  
\mathcal{A}^{yyzz}({\bm X_y \!\otimes\! \bm X_y \!\otimes\!  \bm X_z \!\otimes\! \bm X_z}+ 
{\bm X_y \!\otimes\! \bm X_z \!\otimes\!  \bm X_y \!\otimes\! \bm X_z}
\nonumber\\
&+{\bm X_y \!\otimes\! \bm X_z \!\otimes\!  \bm X_z \!\otimes\! \bm X_y}+{\bm X_z \!\otimes\! \bm X_y \!\otimes\!  \bm X_z \!\otimes\! \bm X_y} 
\nonumber\\
&+{\bm X_z \!\otimes\! \bm X_y \!\otimes\!  \bm X_y \!\otimes\! \bm X_z}+ 
{\bm X_z \!\otimes\! \bm X_z \!\otimes\!  \bm X_y \!\otimes\! \bm X_y})
\nonumber\\
&+\mathcal{A}^{yyyy}{\bm X_y \!\otimes\! \bm X_y \!\otimes\!  \bm X_y \!\otimes\! \bm X_y}+ \mathcal{A}^{zzzz}{\bm X_z \!\otimes\! \bm X_z \!\otimes\!  \bm X_z \!\otimes\! \bm X_z}
\nonumber\\
&+\mathcal{C}^{yyxx}{\bm X_y \!\otimes\! \bm X_y \!\otimes\!  \bm R_x \!\otimes\! \bm R_x}+ \mathcal{C}^{zzxx}{\bm X_z \!\otimes\! \bm X_z \!\otimes\!  \bm R_x \!\otimes\! \bm R_x} 
\nonumber\\
&+
 \mathcal{E}^{xxxx}{\bm R_x \!\otimes\! \bm R_x \!\otimes\!  \bm R_x \!\otimes\! \bm R_x}
\end{align}
The above expression gives the most general rank-4 Killing-St\"ackel tensor  in $\mathbb{E}^2$  consistent with the symmetry  
$(\mathbb{R},+)\times \mathbb{Z}_2 \times \mathbb{Z}_2$.  For clarity, 
let us set define  new parameters  $\kappa,\lambda,\mu,a,b,c$
such that $\mathcal{E}^{xxxx}=\kappa$, $\mathcal{A}^{zzzz}=\lambda$,
$\mathcal{C}^{zzxx}=\mu$, $\mathcal{A}^{yyzz}=(2 \lambda -\mu c^2)/6$, 
$\mathcal{A}^{yyyy}=\lambda-\mu c^2+\kappa a^2 b^2$,
 $\mathcal{C}^{yyxx}=\mu-\kappa(a^2+b^2)$. This reparametrization involves no loss of generality, since $a,b,c$ are allowed to be real or imaginary. Then, the  solution \eqref{eq:Killing2DRank4SolutionSimple} takes the simple form 
\begin{equation} \label{eq:KillingTensorRankFourFull}
{K^{ijkl}} = \kappa A^{(ij}B^{kl)} +\lambda g^{(ij} g^{kl)} + \mu g^{(ij} C^{kl)}
\end{equation}
where the tensors $A^{ij},B^{ij},C^{ij}$ are given by Eqs. 
\eqref{eq:KijAxisymReflReduced} and 
\eqref{eq:KijAxisymReflReducedQart}. 
When contracted with the momenta, the  term 
$g^{(ij} g^{kl)}$ in the above expression is the quartic part of the squared energy, which is conserved; this term is thus dropped from Eq. \eqref{eq:KillingTensorRankFour}.

\bsp

\label{lastpage}

\end{document}